\pdfoutput=1
\documentclass[pra, 10pt, twocolumn, floafix, superscriptaddress] {revtex4-1}

\usepackage{times, mathrsfs, amsmath, amsfonts, graphics, graphicx, cancel, color, theorem, bbm, mathtools, amssymb, soul}

\usepackage[english]{babel}

\usepackage[unicode=true, breaklinks=false, pdfborder={0 0 1}, backref=false, colorlinks=true, linkcolor=blue, citecolor=blue]{hyperref}

\def\bra#1{\mathinner{\langle{#1}|}}
\def\ket#1{\mathinner{|{#1}\rangle}}

\def\BraVert{\egroup\,\mid\,\bgroup}

\def\amplitude#1#2{\mathinner{\langle#1|#2\rangle}}

\newtheorem{definition}{Definition}

\newtheorem{lemma}[definition]{Lemma}

\newtheorem{theorem}[definition]{Theorem}

\begin{document} 

\title{Non-Markovian quantum processes: complete framework and efficient characterisation}

\author{Felix A. Pollock}
\email{felix.pollock@monash.edu}
\affiliation{School of Physics \& Astronomy, Monash University, Clayton, Victoria 3800, Australia}

\author{C\'esar Rodr\'iguez-Rosario}
\affiliation{Bremen Center for Computational Materials Science, University of Bremen, Am Fallturm 1, D-28359, Bremen, Germany}

\author{Thomas Frauenheim}
\affiliation{Bremen Center for Computational Materials Science, University of Bremen, Am Fallturm 1, D-28359, Bremen, Germany}

\author{Mauro Paternostro}
\affiliation{School of Mathematics and Physics, Queen’s University, Belfast BT7 1NN, United Kingdom}

\author{Kavan Modi}
\email{kavan.modi@monash.edu}
\affiliation{School of Physics \& Astronomy, Monash University, Clayton, Victoria 3800, Australia}

\date{\today}

\begin{abstract}

Currently, there is no systematic way to describe a quantum process with memory solely in terms of experimentally accessible quantities. However, recent technological advances mean we have control over systems at scales where memory effects are non-negligible. The lack of such an operational description has hindered advances in understanding physical, chemical and biological processes, where often unjustified theoretical assumptions are made to render a dynamical description tractable. This has led to theories plagued with unphysical results and no consensus on what a quantum Markov (memoryless) process is. Here, we develop a universal framework to characterise arbitrary non-Markovian quantum processes. We show how a multi-time non-Markovian process can be reconstructed experimentally, and that it has a natural representation as a many body quantum state, where temporal correlations are mapped to spatial ones. Moreover, this state is expected to have an efficient matrix product operator form in many cases. Our framework constitutes a systematic tool for the effective description of memory-bearing open-system evolutions. 
\end{abstract}
\maketitle 

\section{Motivation}

No system is isolated. Within its broadest definition, the open systems paradigm embraces this reality and makes use of statistical methods and approximations to account for unknown and uncontrollable variables. It has had tremendous success in translating fundamental theories into real-world predictions and has led to a multitude of technological advances. In quantum mechanics, the conventional description of open dynamics constitutes a mapping from one state of a system to another. However, this approach has serious shortcomings when it comes to describing many realistic scenarios, which has hindered progress in describing complex quantum processes. The reason for these shortcomings is aptly summed up in the famous quote by Asher Peres~\cite{peres}: ``The simple and obvious truth is that quantum phenomena do not occur in a Hilbert space. They occur in a laboratory. If you visit a real laboratory, you will never find there Hermitian operators. All you can see are emitters (lasers, ion guns, synchrotrons and the like) and detectors. The experimenter controls the emission process and observes detection events." In this Article, we embrace Peres's point of view, and propose a new way to describe arbitrary quantum processes in terms of control operations, as opposed to mappings from density operators to density operators. In particular, our framework is perfectly suited to describe temporally correlated, that is \textit{non-Markovian}, quantum processes.

\begin{figure}[t]
\setlength\fboxsep{-1pt}
\begin{center}
\includegraphics[width=1\linewidth]
{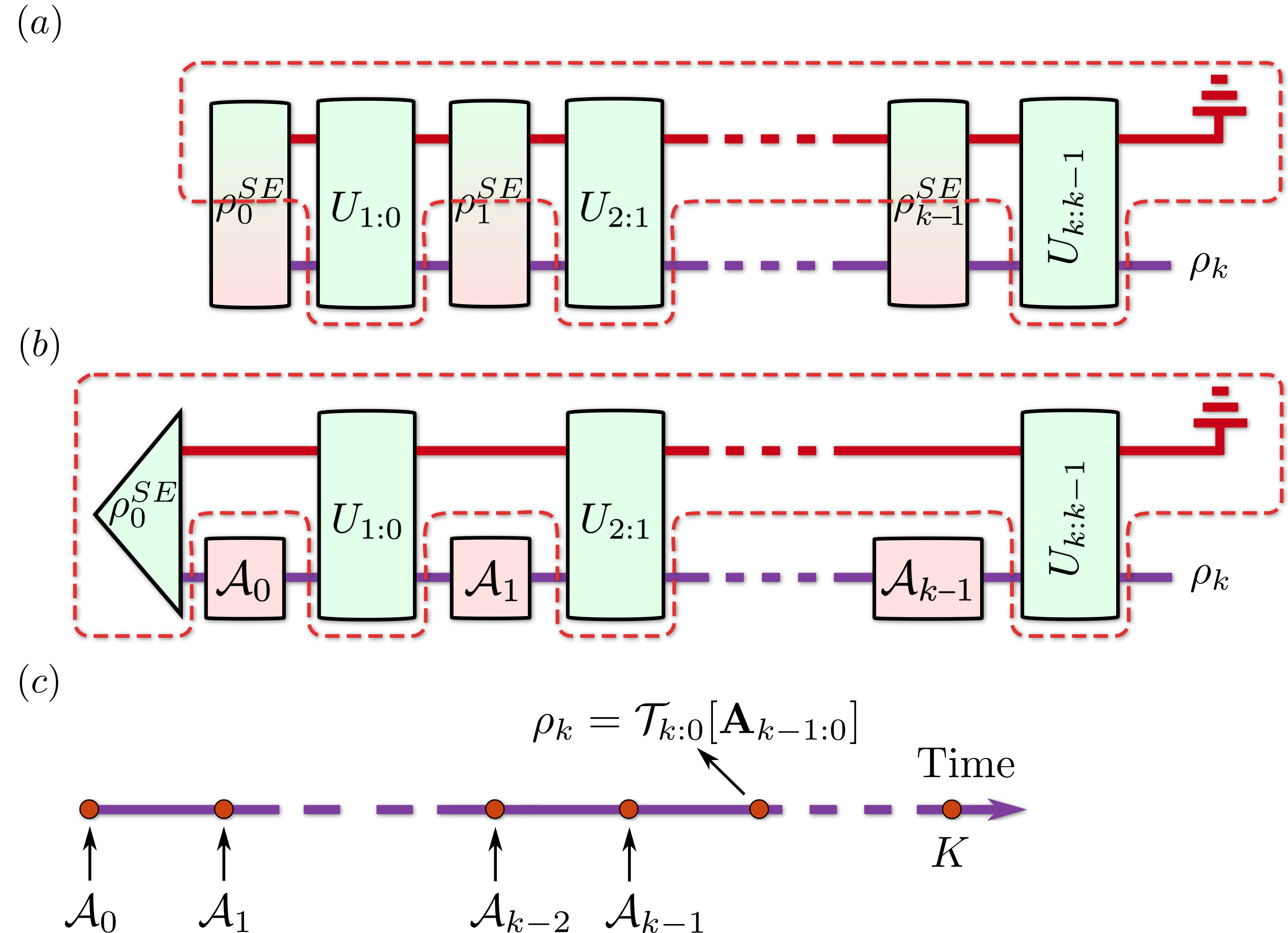}
\caption{{\textbf{(a)} The \textit{conventional approach} to open quantum dynamics attempts to relate the states of the system (${S}$) at different times by considering system-environment (${S\mbox{-}E}$) unitary dynamics and averaging out the state of the environment $({E})$. The averaging of the unknown variables is denoted by the red line. This cuts through ${S\mbox{-}E}$ correlations, leading to issues that have hindered progress in understanding and characterising non-Markovian dynamics. \textbf{(b)} The \textit{operational framework} relates the operations an experimentalist can perform on ${S}$, denoted by $\mathbf{A}_{k-1:0} = \{\mathcal{A}_{k-1} \dots \mathcal{A}_{0} \}$, to the state of ${S}$ at a later time. The red line here cuts between the objects the experimentalist can control and those that they cannot. \textbf{(c)} This leads to the description of a \textit{quantum stochastic process} as a mapping, encapsulated in the process tensor $\mathcal{T}_{k:0}$, from the set of control operations to the output state $\rho_k$ of ${S}$. The process tensor contains all the information about the ${S\mbox{-}E}$ initial state and interactions that can be inferred from the system's dynamics alone.\label{image-q-a}}}
\end{center}
\end{figure}

Future quantum technologies, from quantum computers~\cite{13preskill, 11kalai11060485} to artificial nanostructures~\cite{Lambert:2013xi},
will have to embrace non-Markovian dynamical effects if they are to operate under realistic conditions. Our understanding of fundamental processes in nature, such as the dynamics of molecules~\cite{Nitzan1384} and the functions of bio-chemical systems~\cite{Lambert:2013xi}, also hinges on a clear theory of non-Markovian quantum processes. Already, there are many interesting physical scenarios where going beyond the Markov assumption can be advantageous~\cite{Erez:2008lp}. In such instances, the characterisation of the ensuing dynamics via conventional methods poses many challenges; one often has to relinquish either the \textit{complete positivity} or the \textit{linearity} of the dynamics~\cite{pechukas, alicki, pechukas2}, leading to a mathematically consistent, but physically inapplicable description for the dynamics~\cite{arXiv:1708.00769}---see Fig.~\ref{image-q-a}(a)~\footnote{The states in our figures are density operators, and the operations are superoperators which act on the space of density operators.}. To overcome these difficulties, one must consider that the environment $(E)$, as well as the system-environment $(S\mbox{-}E)$ correlations, might have some memory of previous states of the system $(S)$, significantly complicating any theoretical description~\cite{BreuerPetruccione}. This is particularly true in the quantum regime, where the timescales of the interaction between ${S}$ and ${E}$ are often comparable to those of the dynamics of the system alone~\cite{PhysRevLett.82.2417}. 

In this Article, we present a general \textit{operational} framework to characterise arbitrary quantum processes, including those which are non-Markovian. Our framework closely resembles the quantum combs programme~\cite{PhysRevLett.101.060401, PhysRevA.80.022339} developed to understand the most general quantum circuits. In our framework, a quantum process is defined by the relationship between experimentally implementable controls and experimentally measurable output states, see Fig.~\ref{image-q-a}(b). Our approach is very much in the spirit of Peres's quote above. There are two main results presented in this paper:

\textbf{(I)} A mapping, which we call the \emph{process tensor}, from the set of possible control operations to output states, see Fig.~\ref{image-q-a}(c). We show this mapping is universal, by proving that it describes all quantum processes and can be simulated with a quantum circuit. Our framework is free of any assumptions about the underlying system-environment dynamics, and, unlike many conventional methods in open dynamics, the process tensor naturally accounts for multi-time correlations. We detail the mathematical structure of the process tensor, showing that it retains both linearity and complete positivity, before showing how it can be tomographically reconstructed.

\textbf{(II)} A representation for the process tensor as a many-body quantum state, which can be physically constructed using a set of bipartite entangled states. This many-body state encodes temporal correlations as spatial ones, and has a natural matrix-product-operator representation~\cite{MPS1}. As such, it can be efficiently reconstructed using tensor network techniques developed in recent years. This is our most significant contribution, as it enables an efficient and systematic way to describe non-Markovian quantum processes, and opens the door for the wide range of tools for characterising spatial correlations (e.g. entanglement) to be directly applied to temporal correlations.

Our framework single-handedly resolves the troubling issues surrounding complete positivity and linearity (or lack thereof) faced by the conventional framework when dealing with initial correlations and memory effects. It leads to a complete formulation of open quantum dynamics, in the sense that it describes everything that could possibly be observed in an experiment. Moreover, it could be used to better understand quantum processes such exciton transport, chemical reactions, and many more. It opens up the possibilities for systematically developing techniques for quantum control which will be instrumental in the development of new quantum technologies. In an accompanying Letter, we also use our framework to derive an operationally meaningful Markov condition and corresponding family of measures for non-Markovianity~\cite{PRL_partner}.

\section{Open quantum processes}

\paragraph*{Operational framework}
We consider a quantum system undergoing a process that we split into arbitrary discrete time steps, labelled by $k \in [0,K]$, where we do not assume anything about the intermediate dynamics; nor do we assume anything about the system's initial state, which is a feature of the process itself~\footnote{Every time an experimenter runs the process, the system is in some initial state outside their control (possibly correlated with its environment). They can prepare the system in a new state, but this is an active transformation described by a control operation $\mathcal{A}$.}. When the time steps are chosen to be closely spaced, they will approximate a continuous-time evolution. Within this setting, we begin by giving an operational definition of process characterisation:

\begin{definition}\label{fullchar}
A quantum process is said to be characterised for $K$ time steps when the state of the system can be predicted at any time step $0 \!\! \le k \!\! < K$. The system may be subjected to arbitrary quantum operations $\mathcal{A}$ at previous time steps. The mapping from the sequence of operations $\mathbf{A}_{k-1;0} := \{\mathcal{A}_{k-1}; \dots; \mathcal{A}_1; \mathcal{A}_0\}$ to the state $\rho_k$, given by 
\begin{gather}\label{totalmap}
\rho_k := \mathcal{T}_{k:0}[\mathbf{A}_{k-1:0}],
\end{gather}
fully characterises the process. We call $\mathcal{T}_{k:0}$ the process tensor.
\end{definition}

We have graphically illustrated Definition~\ref{fullchar} in Figure~\ref{image-q-a}(c). This definition of the process tensor, which also encodes the average initial state of the system, forms the basic building block of this work. Operations $\mathbf{A}$ (where we have omitted the subscripts) are called \emph{control operations}: they represent all the possible manipulations of the system -- measurements, unitary rotations etc. --  that an experimentalist could perform, and are mathematically described by \textit{completely-positive} (\textsc{cp}) maps. When the operations can be performed deterministically (for example, a unitary rotation), they are also trace preserving \textsc{cptp} maps. Otherwise, when a control can only be applied probabilistically, corresponding to a particular measurement outcome for instance, the trace of the state is decreased. In this case, the output of the process tensor is a subnormalised density matrix proportional to the success probability of applying the trace decreasing controls.

In general, the control operations may even be correlated with one another, corresponding to classical conditioning or multiple interactions with the same ancillary system. Their only restriction is that they must act on ${S}$ alone. An important subset of control operations is the combination of a measurement followed by a preparation. Definition~\ref{fullchar} represents the idea that an experimentalist can probe a system  many times, and in many different ways, as it evolves, and that the full statistics of all possible observations constitutes the effective process accessible to the experimentalist.

\subsection{Properties of the process tensor}
\label{sec:props}

The process tensor is a mapping from the set $\mathbf{A}$ to a quantum state. Thus, its output is required to be a valid density operator, up to normalisation (which depends on the probability of applying $\mathbf{A}$). Furthermore, it should satisfy the following properties to be physically relevant:
\begin{enumerate}
\item [{\bf(P1)}] \emph{Linearity:} $\mathcal{T} [a \mathbf{A} + b \mathbf{B}] = a \mathcal{T} [ \mathbf{A}] + b \mathcal{T} [\mathbf{B}]$ for any $a,b \in \mathbb{R}$. This property embodies the linearity of mixing, which must hold for any stochastic theory.

\item[{\bf(P2)}] \emph{Complete positivity:} If the controls act on the system ${S}$ undergoing the process and an ancilla ${A}$, the final $S$-$A$ state should still be physical. Therefore $\mathcal{T}^{S} \otimes \mathcal{I}^{A} [\mathbf{A}^{{SA}}] = \rho^{{SA}} \ge 0$, where $\mathcal{I}^{A}$ is the identity process on the ancilla; this must be true for any $\mathbf{A}^{{SA}}$. This is analogous to complete positivity for quantum operations.

\item [{\bf(P3)}] \emph{Containment:} For $k \ge k' \ge j' \ge j$, the process tensor $\mathcal{T}_{k':j'}$ is contained in $\mathcal{T}_{k:j}$. That is, if we have the full process tensor $\mathcal{T}_{K:0}$, then we can describe the dynamics between any intermediate time steps, and $\mathcal{T}_{k':j'}$ can be obtained from $\mathcal{T}_{k:j}$. This amounts to a causal ordering of time steps.
\end{enumerate}

We now prove that the process tensor given in Definition~\ref{fullchar} with these properties fully describes \emph{any} quantum process -- even when it involves strong system-environment coupling -- and is guaranteed to have physical outputs. Unlike conventional approaches, the process tensor has all of the desired properties of a statistical-dynamical theory -- linearity, a notion of complete positivity etc. -- while accounting for arbitrary non-Markovian behaviour.

\subsection{Representation theorem}
\label{sec:repthm}


We use the term open quantum evolution (OQE) to describe a system ${S}$ interacting with its environment ${E}$, where the joint ${S\mbox{-}E}$ dynamics is driven by unitary evolution, i.e., according to the Schr\"odinger equation. As above, the system may be interrogated, interrupted, or manipulated at intermediary time steps by controls $\mathbf{A}_{k-1:0} = \{\mathcal{A}_{k-1} \dots \mathcal{A}_0\}$, which are simply \textsc{cp}~operations.

We can write the total dynamics as
\begin{gather}\label{eq:oqe}
\rho^{SE}_k := \mathcal{U}_{k:k-1} \, \mathcal{A}_{k-1} \, \mathcal{U}_{k-1:k-2} \dots \mathcal{A}_1 \, \mathcal{U}_{1:0} \, \mathcal{A}_0 \, [\rho^{SE}_{0}],
\end{gather}
where $\rho^{SE}_{0}$ is the initial ${S\mbox{-}E}$ state, $\{\mathcal{U}\}$ are unitary maps on the ${S\mbox{-}E}$ space given by $\mathcal{U}_{j:i} [\rho^{SE}_{i}] = U_{j:i} \rho^{SE}_{i} \, U^\dagger_{j:i} = \rho^{SE}_{j}$, where $U_{j:i}U^\dagger_{j:i}=\mathbbm{1}$ and $\rho^{SE}_k$ is the state of ${S\mbox{-}E}$ at time step $k$. The state of the system is obtained by tracing over the environment as $\rho^{S}_k = {\rm tr}_{E}[\rho^{SE}_k]$. Equation~\ref{eq:oqe} is the full quantum mechanical description of the joint ${S\mbox{-}E}$ evolution. We now formalise the relationship between the process tensor and OQE with the following Theorem.
\begin{theorem}
\label{thm:rep}
The state of a system, undergoing an open quantum evolution, at any time step $k$ is given by contracting a choice of control operations with a process tensor satisfying the properties: \textbf{(P1)} linearity; \textbf{(P2)} complete positivity; and \textbf{(P3)} containment. Conversely, any process tensor is consistent with an OQE of the form of Eq.~\eqref{eq:oqe}, where the environment is simulated by $k$ ancillas of increasing dimension $d_{{A}_j}\ge d^{2 (3^j)}$.
\label{thm:dilation}
\end{theorem}
The proof of the first statement, given in Appendix~\ref{app:proof:thm:qdptopt}, constructs $\mathcal{T}_{k:0}$ explicitly by writing down the matrix indices for all objects in Eq.~\eqref{eq:oqe}. Specifically, to prove the Theorem, we show that the action of the process tensor can be written as the operator-sum decomposition
\begin{gather}
\rho = \mathcal{T} [\mathbf{A}] = \sum_l T_l \, \mathbf{A} \, T_l^\dag, \label{eq:cpform}
\end{gather}
with the operators $\{T_l\}$ defined in Eq.~\eqref{krausform}. The second equality implies complete positivity (and linearity) of $\mathcal{T}$. The containment property also arises naturally from our construction. 

For the proof of the converse statement, given in Appendix~\ref{app:dilationproof}, we make use of the supermaps formalism introduced in Ref.~\cite{supermaps}. In a nutshell, we show that each step of a process can be described by a supermap, and that this implies a unitary representation for the dynamics during that step. By induction, the unitary representation, or dilation, of the full process tensor follows. It is also possible to represent a general process tensor by a unitary evolution with ancillas of smaller dimension $d_{A_j}\geq d^{2k+1}$~\cite{combsdilation}, albeit with a circuit that cannot be straightforwardly extended to incorporate more time steps.

\begin{figure}[t]
\begin{center}
\includegraphics[width=.9 \linewidth] {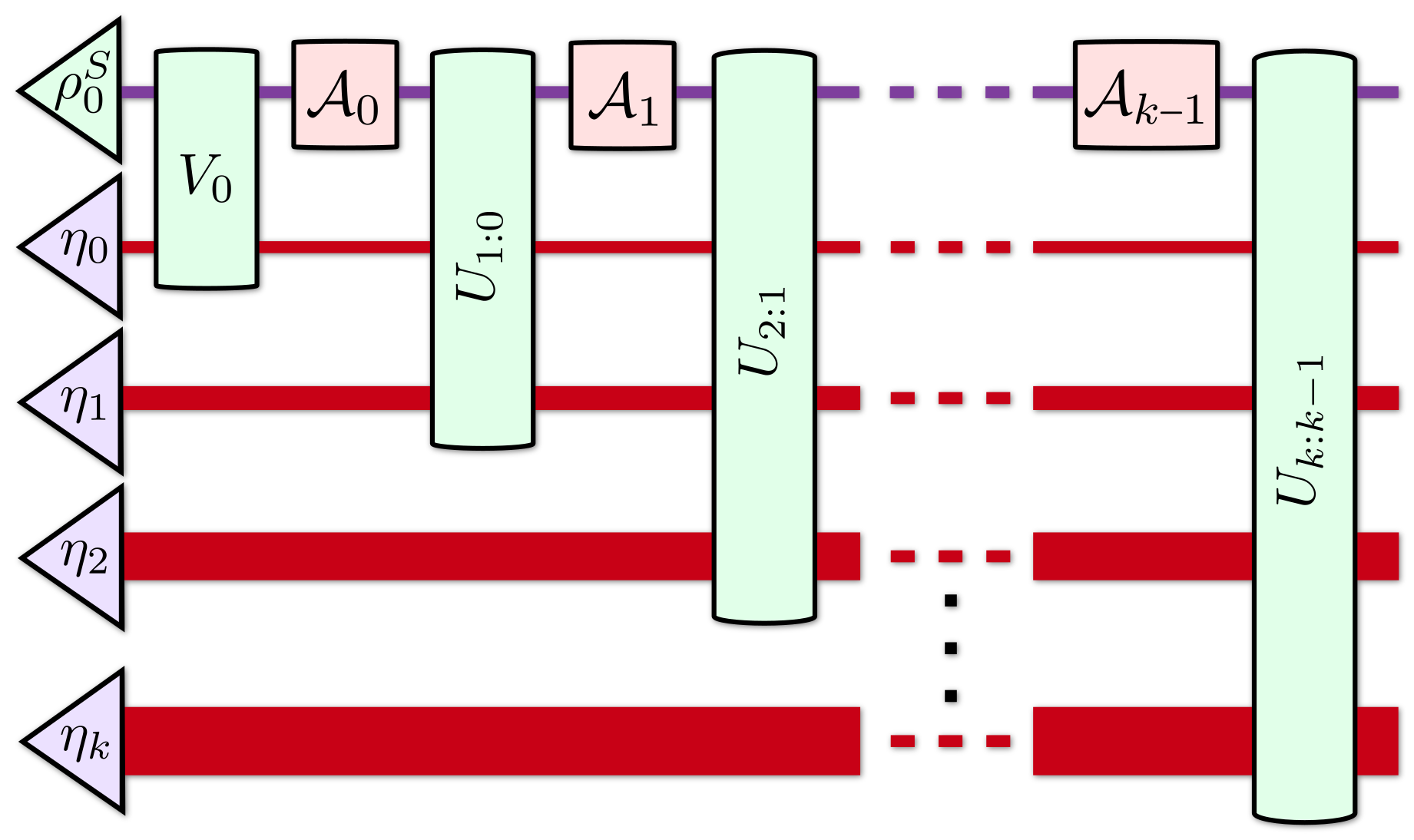}
\caption{\emph{Quantum circuit to simulate the process tensor.} Any process tensor can be simulated by the quantum circuit above. For each time step an ancilla of dimension $d_j\ge(d_S\prod_{n=0}^{j-1} d_n)^2$ and prepared in a state $\eta_j$ is introduced. The unitary at each step can be decomposed as $U_{j:j-1} = V_j W_{j:j-1}$, where $W_{j:j-1}$ acts on the system and all previous ancilla and $V_j$ acts on all subsystems including the new ancilla. See Appendix~\ref{app:dilationproof} for a detailed proof of {the converse statement of Theorem}~\ref{thm:rep}.} \label{fig:dilateddynamics}
\end{center}
\end{figure}

The Theorem above show that the process tensor is the most general descriptor for a quantum process. The direct correspondence between OQE and the process tensor proves its universality. It additionally provides a recipe for simulating general (discrete-time) non-Markovian dynamics. Given a process tensor description of the dynamics, a set of unitary operations $\{\mathcal{U}_{j:j-1}\}$ and ancilla states $\{\eta_j\}$ can be (non-uniquely) determined which, when applied using the quantum circuit in Figure~\ref{fig:dilateddynamics}, fully simulate the reduced dynamics of the system. Since this description is operational, it is experimentally applicable, sidestepping issues of interpretation of all other approaches. Nevertheless, since it can describe any quantum process, the process tensor contains the conventional picture in the latter's realm of validity.

\subsection{Conventional picture from the process tensor}
\label{app:conv}


In the conventional approach the dynamics of a quantum system is most often described by a master equation or a family of dynamical maps. The former relates the rate of change of a system's state (represented by a density operator) at each time to the state itself, or more generally to the state at earlier times. On the other hand, in the latter approach, the future states of the system are obtained by the action of a superoperator on the initial state. In other words, the conventional description of a process involves keeping track of the state of the system as a function of time. This limits the ability to determine the outcomes of measurements on the system to at most two times in a given run, failing to account for multi-time correlations, which are crucial for understanding non-Markovian effects. Moreover, the conventional approach runs into trouble when when the initial state of $S\mbox{-}E$ is correlated.

The presence of initial $S\mbox{-}E$ correlations indicates one of the simplest non-Markovian processes; the initial correlations are a record of the past interactions between $S$ and $E$. In such cases, the \textsc{cptp}~description of the dynamics breaks down. Pechukas has shown that, in order to describe the dynamics in the presence of initial $S\mbox{-}E$ correlations, we must give up something~\cite{pechukas, alicki}, e.g. complete positivity or linearity~\cite{pechukas2}. Needless to say, neither of these two options is desirable, creating a double-bind. The operational interpretation of non-\textsc{cp}~or nonlinear maps is not clear, and they can lead to unphysical behaviour~\cite{kuah, PhysRevA.81.012313, modi-sudarshan}. These troubling features remain when describing general (and more complex) non-Markovian dynamics~\cite{Rodriguez11b, NMrev, breuer-rev}.

To overcome the double-bind presented by the initial correlation problem there is a a third option: to give up altogether the notion of states of $S$ as the inputs of the map~\cite{modiscirep}. This is because an independent set of input states of $S$ is not well defined when $S$ is correlated with $E$~\cite{modiosid}. If we recognize that in order to prepare a desired state of $S$, we must, in reality, implement some external control operation, then, it is thus natural to treat these operations as the inputs to the process~\cite{modiscirep}, which in turn yields the final state of $S$. This method is an operationally sound way of describing dynamics when the initial $S\mbox{-}E$ state is correlated, and has been experimentally implemented~\cite{martin}. The resultant map is a single step process tensor, also known as a \emph{superchannel}. As such,  it is both \textsc{cp}~and linear, overcoming the challenge posed by Pechukas.

While the superchannel resolves the problem of initial correlations, the more general process tensor allows for describing correlations over multiple times steps. It too maps control operations to states, instead of initial system states to final states, and it is fundamentally different from the conventional approach to non-Markovian dynamics~\cite{PhysRevLett.101.150402, PhysRevLett.103.210401, huelga, Rodriguez11b, nicola, sabri}. It also differs from non-Markovian master equations, which seek to relate changes in the state of a system at a given time to its initial state and the effects of a memory kernel; this may be microscopically derived or phenomenological in nature~\cite{BreuerPetruccione, nmqsd, TCLPO}.

\begin{figure}
\begin{center}
\includegraphics[width=1.0 \linewidth]{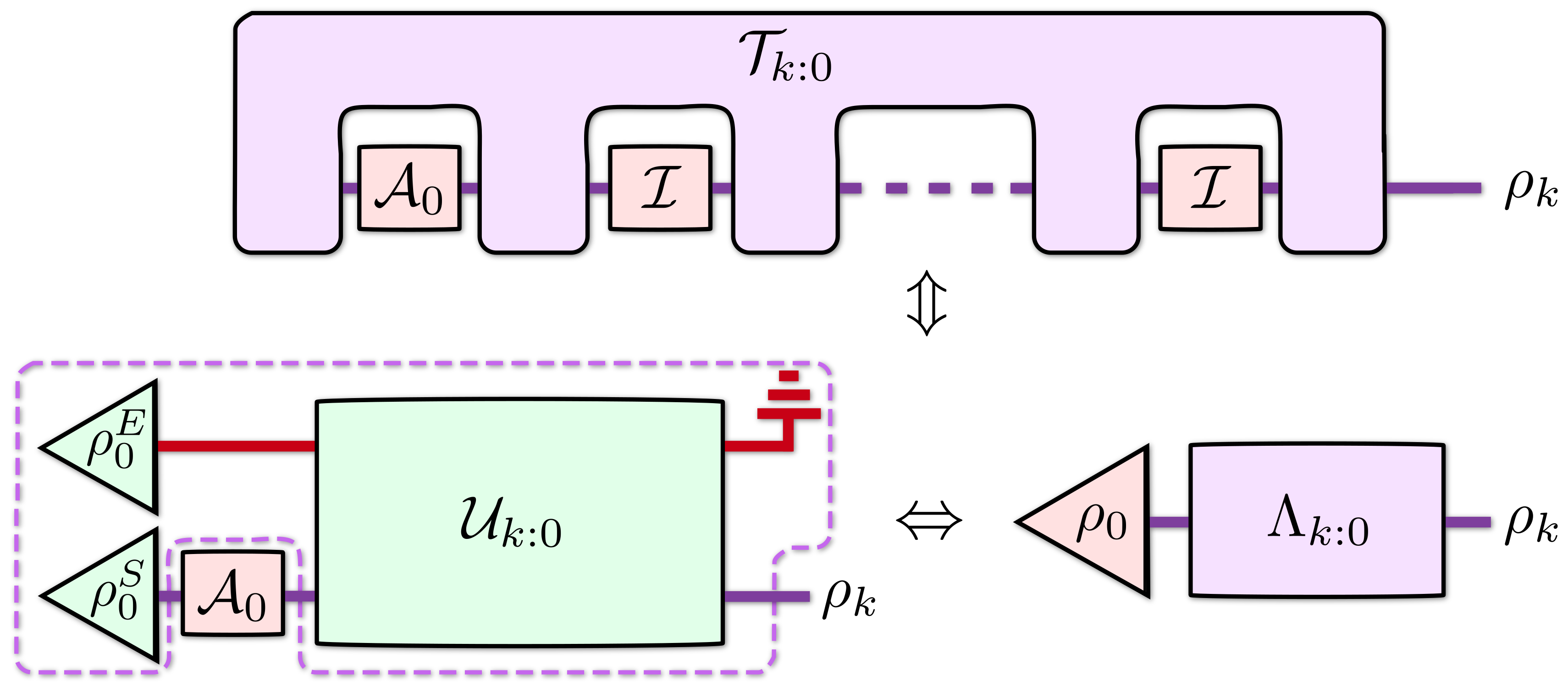}
\caption{\emph{The conventional picture of open dynamics} is fully contained in the process tensor. When the initial system-environment state is uncorrelated (i.e., $\rho_0^{SE} = \rho_0^S \otimes \rho_0^E$), the picture of evolution according to a \textsc{cptp} map can be recovered by acting with the identity map $\mathcal{I}$ (doing nothing) at all time steps but the first. The initial state is simply given by $\rho_0=\mathcal{A}_0[\rho_0^S]$.\label{fig:conventional}}
\end{center}
\vspace{-20pt}
\end{figure}

As it is a more general description, the process tensor includes the same information (and more) about the dynamics as the conventional approach. In particular, it can be used to determine the density operator as a function of time. Let us assume that the initial $S\mbox{-}E$ state is uncorrelated. Thus, the state of $S$ at time step $k$ is given by
\begin{gather}
\rho_k = \mbox{tr}_E [U_{k:0} \ \rho^S_0 \otimes \rho^E_0 \ U^\dag_{k:0}] = \Lambda_{k:0}(\rho_0),
\end{gather}
where $\Lambda_{k:0}$ is a \textsc{cptp} map from the initial time to time step $k$. This expression can be obtained from the process tensor by simply choosing the identity operation (do nothing) as the control operation at each time step after the initial preparation:
\begin{gather}
\rho_k = \mathcal{T}_{k:0}[\mathcal{I};\dots;\mathcal{I};\mathcal{A}_0].
\end{gather}
This equivalence is depicted in Fig.~\ref{fig:conventional}.

Moreover, by taking time steps closer and closer together, we can also recover the changes to the state of the system. This allows for deriving a non-Markovian master equation of the Nakajima-Zwanzig type~\cite{PollockModi2017}. While the conventional approaches are recovered as limiting cases, the process tensor allows for much more, including implementing temporally correlated control operations.

\subsection{Temporally correlated controls}
\label{app:timecorrelatedops}

The linearity property (\textbf{P1}) of the process tensor applies independently to each of its arguments, that is the process tensor is multi-linear in the applied control operations: 
\begin{align}
&\mathcal{T}_{k:0}[\mathcal{A}_{k-1}; \dots; (a\mathcal{A}_j+b\mathcal{A}_j') ; \dots; \mathcal{A}_0] = \nonumber\\ 
&\qquad \qquad \qquad \qquad a\mathcal{T}_{k:0}[\{\mathcal{A}_{k-1}; \dots; \mathcal{A}_j; \dots;  \mathcal{A}_0\}] \\ \nonumber
&\quad \qquad \qquad \qquad \qquad +b\mathcal{T}_{k:0}[\{\mathcal{A}_{k-1}; \dots; \mathcal{A}_j'; \dots;  \mathcal{A}_0\}]
\end{align}
$\forall j \in [0,k-1]$ and 
$\forall a,b \in \mathbb{R}$. What this means is that the argument $\mathbf{A}=\{\mathcal{A}_{k-1}; \dots; \mathcal{A}_1; \mathcal{A}_0\}$ can be seen as an element of the tensor product space of control operations. For independent operations this means we can write $\mathbf{A}=\mathcal{A}_{k-1}\otimes \dots\otimes\mathcal{A}_1\otimes\mathcal{A}_0$.

\begin{figure}
\begin{center}
\includegraphics[width=1.0 \linewidth]{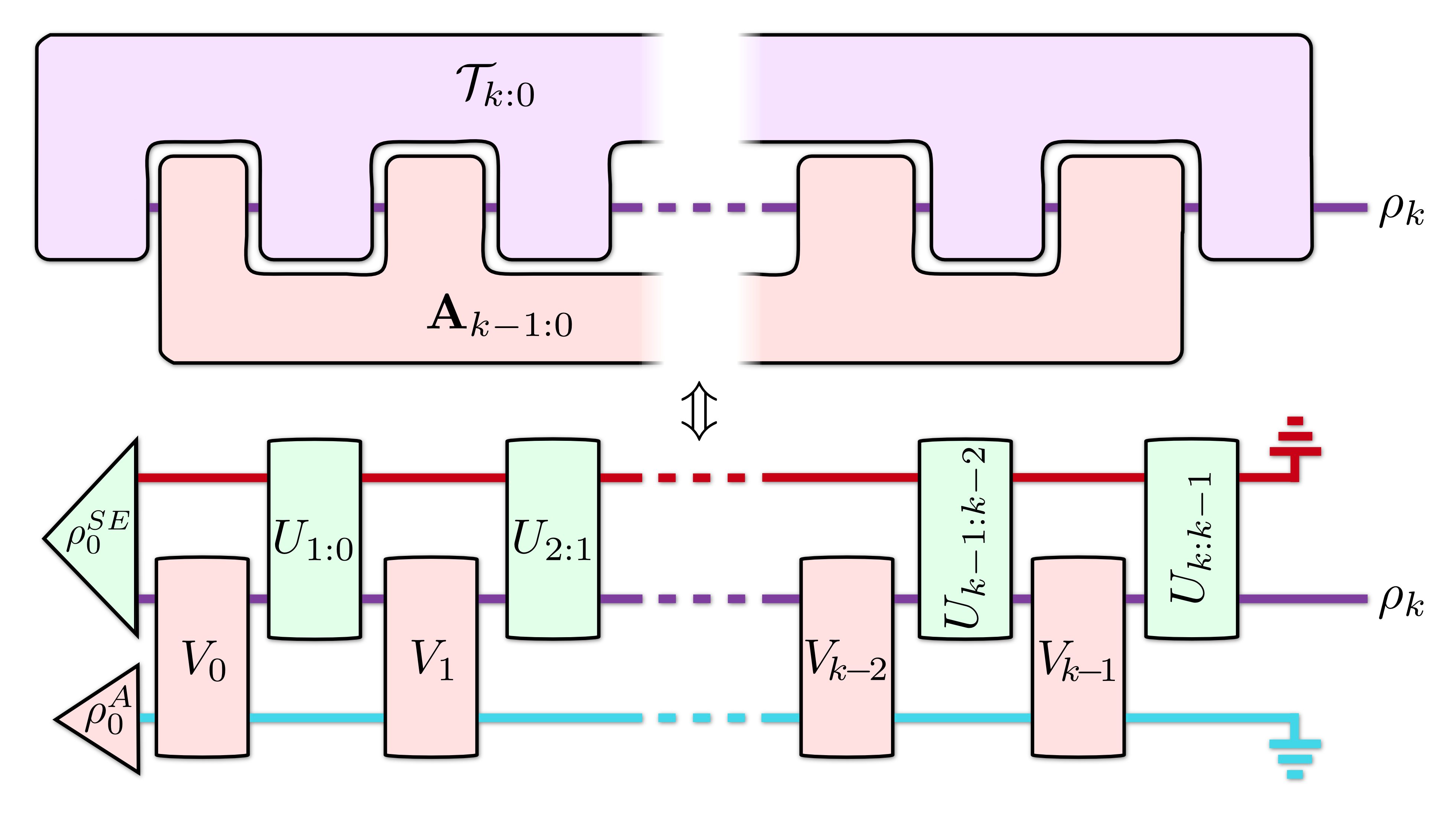}
\caption{\emph{A process tensor acting on a correlated operation.} Both the process tensor and the sequence of control operations can be represented as quantum combs~\cite{PhysRevA.80.022339}. The density operator at time step $k$ results from their contraction.  Any correlated \textsc{cptp}~operation acting on the system can be implemented by interacting the system repeatedly with an ancillary system ${A}$, such that $SA$ unitaries $V_j$ are applied at each time step. Any correlated \textsc{cp} operation can be implemented with a further measurement with the correct outcome~\cite{Peres1990}.}\label{fig:comboncomb}
\end{center}
\vspace{-20pt}
\end{figure}

Noting the tensor product structure of the process tensor's argument means that we can extend its action to non-product operations 
\begin{gather}
\mathbf{A}=\sum_{j_{k-1},\dots,j_0}c_{j_{k-1},\dots,j_1,j_0}\mathcal{A}_{j_{k-1}}\otimes \dots\otimes
\mathcal{A}_{0}.    
\end{gather}
These correspond to correlated operations; for example, these could be measurements whose basis depends on the outcome of an earlier measurement, or they could represent repeated interactions with the same ancillary system. In fact, these operations also have the structure of a general quantum comb, and hence can be thought of as process tensors themselves (with an uncorrelated initial state). In Fig.~\ref{fig:comboncomb} we depict the action of the process tensor on a general correlated operation, and how this could be realised in practice. Correlated operations can be used to describe experiments with quantum or classical feedback control. In the following section, we further use the linearity of the process tensor and the control operations to show how it can be reconstructed tomographically.

\section{Linear construction of the process tensor}
\label{app:construction}
The collection of  quantum operations  $\mathbf{A}$, which the process tensor acts on, is itself a linear operation. That is, an operation $\mathcal{A}$, acting on the system at a given time step, is a linear map on the density operator of the system. At each time step $j$, it can be uniquely decomposed in terms of a fixed set of linearly independent operations $\{\mathcal{A}_j^{{(\mu,\nu)}_j}\}$ as $\mathcal{A}_j = \sum_{{(\mu,\nu)}} \alpha_{{(\mu,\nu)}_j} \mathcal{A}_j^{{(\mu,\nu)}_j}$, with real numbers $\alpha_{{(\mu,\nu)}_j}$.
Note that the coefficients $\alpha_{{(\mu,\nu)}_j}$ are not necessarily positive, meaning the expansion above is linear but not convex. Further, using the multilinearity of the process tensor discussed in the previous section, any sequence of control operations can also be expanded in terms of tensor products of these basis elements as
\begin{gather}\label{MesPrep}
\mathbf{A}_{k-1:0} = \sum_{{(\mu,\nu)}} \bigotimes_{j=0}^{k-1} \alpha_{{(\mu,\nu)}_j} \mathcal{A}^{{(\mu,\nu)}_j}_j.
\end{gather}

As we will now see, by determining the final state for each basis operation, the process tensor can be reconstructed in a process tomography~\cite{nielsen, poyatos,NielsenBook} involving many time-steps. As with any quantum tomography, the scaling is not favourable. An operation on any $d$-dimensional system can be expressed in terms of $\mathcal{O}(d^2)$ measurement operators and $d^2$ preparations. Thus, the expansion of $\mathbf{A}_{k:0}$ requires $\mathcal{O}(d^{4k})$ linearly independent combinations of preparations and measurements (or, more generally, $\mathcal{O}(d^{4k})$ linearly independent operations of any sort). This may seem like an obstacle in characterising non-Markovian processes. However, it is still possible to tomographically reconstruct a partial process tensor with a smaller set of controls~\cite{arXiv:1610.02152}.

\begin{figure}[t]
\begin{center}
\includegraphics[width=.95\linewidth]
{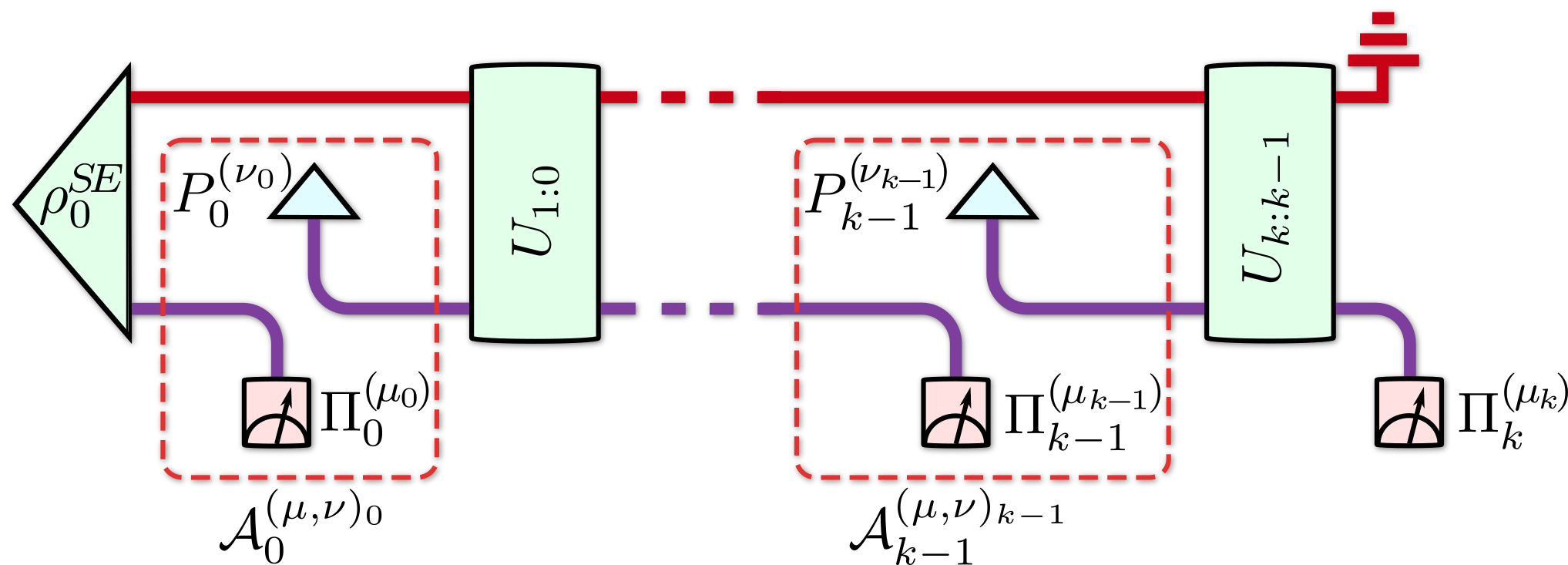}
\caption{\emph{Full-process tomography.} In a convenient, but not unique scheme for full tomography, the system is measured at each time step and then freshly prepared. That is, the preparation at step $k$ is independent of the previous measurements and preparations. A linear combination of measurements and preparations, each chosen from a set that linearly spans the operator space, is sufficient to span the space of control operations. Having statistics for all possible measurements and preparations at all times is sufficient to construct $\mathcal{T}_{k:0}$. 
\label{image-q-op}}
\end{center}
\vspace{-20pt}
\end{figure}

A convenient choice $\{\mathcal{A}^{{(\mu,\nu)}}\}$ for the basis of operations, depicted in Fig.~\ref{image-q-op}, is each of the outcomes of an informationally-complete positive-operator-valued measure (POVM) $\{ \Pi^{(\mu)} \}$ followed by an update~\cite{info-comp}. The update is a preparation of a fresh state from the set $\{P^{(\nu)}\}$, which linearly spans the space of system density operators: $\mathcal{A}^{{(\mu,\nu)}} (\rho) = P^{(\nu)} {\rm tr}[\Pi^{(\mu)} \rho]$, with $\sum_\mu \Pi^{(\mu)} = \mathbbm{1}$. For convenience, we can write the operations in terms of their Choi state~\cite{arXiv:1708.00769} (see also, Sec.~\ref{sec:choi}), which for the basis elements is the simple tensor product $\mathcal{A}^{{(\mu,\nu)}}=P^{(\nu)} \otimes {\Pi^{(\mu)}}$; we will use this representation for the remainder of this section. Preparation of the `fresh' state need not involve another copy of the system, but could be achieved by applying an outcome dependent unitary operation after the measurement (such that the preparation is fully \textit{independent} of the measurement outcome). 

The full control set on a set of time steps can also be cast as a linear combination of sequences of measurements and preparations at each time step. That is, the Choi state of the sequence can be decomposed as
\begin{gather}
\mathbf{A}_{k-1:0}= \sum_{{\vec{\mu}},{\vec{\nu}}}\alpha_{({\vec{\mu}},{\vec{\nu}})}\bigotimes_{j=0}^{k-1} P_j^{(\nu_j)}\otimes\Pi_j^{(\mu_j)},   
\end{gather}
where the notation $\vec{\eta}$ is shorthand for the list of indices $\{\eta_{k-1},\cdots,\eta_1,\eta_0\}$ corresponding to each time step, and we have allowed for the basis $\{\mathcal{A}^{{(\mu,\nu)}_j}_j\}$ to be different at different time steps. When the operations applied at each time step are independent, the coefficients can be decomposed into a product $\alpha_{({\vec{\mu}},{\vec{\nu}})}=\prod_j\alpha_{(\mu,\nu)_j}$. 

Writing the state at time step $k$ as the action of the process tensor on $\mathbf{A}_{k-1:0}$, we can use the above decomposition to express it in terms of a fixed set of basis states:
\begin{widetext}
\begin{align}
\rho_{k}(\mathbf{A}_{k-1:0}) =& \mathcal{T}_{k:0}(\mathbf{A}_{k-1:0}) = \!\!\!\sum_{{(\mu,\nu)}_{k-1}}\!\!\! \cdots\!\!
 \sum_{{(\mu,\nu)}_0} \alpha_{(\vec{\mu},\vec{\nu})} \, \rho_{k} \left(\mathcal{A}_{k-1}^{{(\mu,\nu)}_{k-1}}; \mathcal{A}_{k-2}^{{(\mu,\nu)}_{k-2}}; \dots;\mathcal{A}_1^{{(\mu,\nu)}_{1}}, \mathcal{A}_0^{{(\mu,\nu)}_{0}}\right)\nonumber\\
=&\!\!\! \sum_{{(\mu,\nu)}_{k-1}} \!\!\!\cdots\!\! \sum_{{(\mu,\nu)}_0}\! \alpha_{(\vec{\mu},\vec{\nu})} \,
 \rho_{k} \left(P_{k-1}^{(\nu_{k-1})}, \Pi_{k-1}^{(\mu_{k-1})}
 ; \dots; P_1^{(\mu_{1})}, \Pi_1^{(\mu_{1})}; P_0^{(\mu_{0})}, \Pi_0^{(\mu_{0})} \right).
 \label{linbreak2}
\end{align}
Let us further denote output states for the input basis elements
\begin{gather}
\rho_{k} (\vec{\mu},\vec{\nu}):= \rho_{k} \left(P_{k-1}^{(\nu_{k-1})},\Pi_{k-1}^{(\mu_{k-1})}; P_{k-2}^{(\nu_{k-2})}, \Pi_{k-2}^{(\mu_{k-2})}; \dots; P_1^{(\mu_{1})}, \Pi_1^{(\mu_{1})}; P_0^{(\mu_{0})}, \Pi_0^{(\mu_{0})} \right).
\end{gather}
\end{widetext}
Since the basis elements correspond to non-deterministic operations (particular measurement outcomes), these states are subnormalised. The trace of one of these states gives the joint probability $p_k(\vec{\mu},\vec{\nu})={\rm tr}[\rho_{k}(\vec{\mu},\vec{\nu})]$ to measure the sequence of outcomes corresponding to POVM elements $\{\Pi_j^{(\mu_j)}\}$ given the set of preparations $P_j^{(\nu_j)}$. Quantum state tomography on the system after a given sequence of basis operations would give the normalised conditional state 
\begin{gather} \label{conditionalstate}
\rho_{k} \left(P_{k-1}^{(\nu_{k-1})} \big| \Pi_{k-1}^{(\mu_{k-1})}; \dots; P_0^{(\mu_{0})}, \Pi_0^{(\mu_{0})}\right) = \frac{\rho_{k}(\vec{\mu},\vec{\nu})}{p_{k}(\vec{\mu},\vec{\nu})}.
\end{gather}

Equation~\eqref{linbreak2} tells us that reconstructing the set of states $\rho_{k}(\vec{\mu},\vec{\nu})$ for all possible values of $(\vec{\mu},\vec{\nu})$ is sufficient to construct the state $\rho_k$ for any arbitrary choice of operations $\mathbf{A}_{k-1:0}$. We simply need to know the expansion coefficients for the sequence, i.e., $\alpha_{(\vec{\mu},\vec{\nu})}$. This is a consequence of the linearity of the process tensor: Given a set of operations, spanned by some control parameters, an experimentalist can test which operations are  linearly independent---this is just a more involved version of quantum process tomography. By a linear inversion process, using $\mathcal{A}^{{(\mu,\nu)}_j}$ and $\rho_{k}(\vec{\mu},\vec{\nu})$ we can construct the map $\mathcal{T}_{k:0}$ which fully characterises the process up to time step $k$. Note again that the set of experiments we are prescribing here simply involve performing a POVM $\Pi_k = \{\Pi_k^{(\mu_k)}\}$ followed by an update $P_k= \{P_k^{(\nu_k)}\}$ at each time step. It is important to note that both $\Pi_k$ and $P_k$ only contain a finite number of elements. Performing (exponentially many) experiments with randomised measurements and preparations will sample from all possible combinations. The states $\rho_{k}(\vec{\mu},\vec{\nu})$ are simply deduced from quantum state tomography of the conditional states in Eq.~\eqref{conditionalstate} and the statistics of the $\Pi_k$ while holding all of the priors constant, since the POVM is informationally complete. We now give a Lemma (analogous to the one given in Ref.~\cite{modi2012positivity}) which allows us to construct the process tensor as a matrix.
\begin{lemma}
The process tensor can be constructed as 
\begin{gather}
\mathcal{T}_{k:0} = \sum_{\vec{\mu},\vec{\nu}} \rho_{k}(\vec{\mu},\vec{\nu}) \otimes D_{\vec{\nu}}^T \otimes \Delta_{\vec{\mu}}^T.
\end{gather}	
where $\{ D_{\vec{\nu}} \}$ and $\{ \Delta_{\vec{\mu}} \}$ are the dual matrices to 
$\{P_{\vec{\nu}}\}$ and $\{\Pi_{\vec{\mu}}\}$ satisfying ${\rm tr}[D_{\vec{\nu}'} \; P_{\vec{\nu}}] =\prod_j\delta_{\nu_j \nu'_j}$ and ${\rm tr}[\Delta_{\vec{\mu}'} \; \Pi_{\vec{\mu}}] =\prod_j\delta_{\mu_j \mu_j'}$.
\end{lemma}

\emph{Proof.} We first prove that for any set of linearly independent matrices $\{P^{(\nu)}\}$ there exists the dual set $\{D^{(\nu)}\}$. Write $P^{(\nu)} = \sum_{\nu'} h_{\nu \nu'} \Gamma^{(\nu')}$, where $h_{\nu \nu'}$ are real numbers and $\{\Gamma^{(\nu')}\}$ form a Hermitian self-dual linearly independent basis satisfying ${\rm tr}[\Gamma^{(\nu)} \Gamma^{(\nu')}]=2 \delta_{\nu \nu'}$~\cite{modi2012positivity}. Since $\{P^{(\nu)}\}$ form a linearly independent basis, the columns of matrix $\mathsf{H} = \sum_{\nu \nu'} h_{\nu \nu'} \ket{\nu}\bra{\nu'}$ are linearly independent vectors, which means $\mathsf{H}$ has an inverse. Let matrix $\mathsf{J}^T=\mathsf{H}^{-1}$, then $\mathsf{H} \mathsf{J}^T = \mathbbm{1}$, implying that the columns of $\mathsf{J}$ are orthonormal to the columns of $\mathsf{H}$. We define $D^{(\nu')} = \frac{1}{2} \sum_j d_{\nu \nu'} \Gamma^{(\nu')}$, where $d_{\nu \nu'}$ are elements of $\mathsf{J}$. The same proof applies to $\{\Pi^{(\mu)}\}$, whose dual set is $\{\Delta^{(\mu)}\}$: ${\rm tr}[\Delta^{(\mu')} \; \Pi^{(\mu)}] =\delta_{\nu \nu'}$. Since, $D_{\vec{\nu}}= \bigotimes_j D_j^{(\nu_j)}$, $\Delta_{\vec{\mu}}= \bigotimes_j \Delta_j^{(\mu_j)}$, $P_{\vec{\nu}}= \bigotimes_j P_j^{(\nu_j)}$ and $\Pi_{\vec{\mu}}= \bigotimes_j \Pi_j^{(\mu_j)}$, we have ${\rm tr}[D_{\vec{\nu}'} \; P_{\vec{\nu}}] =\prod_j\delta_{\nu_j \nu'_j}$ and ${\rm tr}[\Delta_{\vec{\mu}'} \; \Pi_{\vec{\mu}}] =\prod_j\delta_{\mu_j \mu_j'}$.

The action of the process tensor on a specific choice $P_{\vec{\nu}} \otimes \Pi_{\vec{\mu}}$ is given as
\begin{align}
\mathcal{T}_{k:0} [P_{\vec{\nu}} \otimes \Pi_{\vec{\mu}}] =& \sum_{\vec{\mu}',\vec{\nu}'} \rho_{k}(\vec{\mu}',\vec{\nu}')
\; {\rm tr} [D_{\vec{\nu}'}\, P_{\vec{\nu}}] 
\; {\rm tr}[\Delta_{\vec{\mu}'} \, \Pi_{\vec{\mu}}] \nonumber \\ =& \rho_{k}(\vec{\mu},\vec{\nu}).
\end{align}
Its action is then defined on any control operation $\mathbf{A}_{k-1:0}$, by linearly expanding the latter in terms of $P_{\vec{\nu}} \otimes \Pi_{\vec{\mu}}$ and coefficients $\{\alpha_{(\vec{\mu},\vec{\nu})}\}$. The above decomposition therefore provides an operational means to construct the process tensor.
\hfill $\blacksquare$

While the process tensor can be reconstructed in a finite number of experiments this way, the complexity of the procedure scales exponentially with the number of time steps. In the following section, we will discuss an alternative representation for the process which can make its description more efficient.

\section{Efficient state representation of the process tensor}
\label{sec:choi}

To efficiently describe a quantum process, we will map the process tensor into a many-body quantum state. For \textsc{cptp}~maps, there is a remarkable relationship known as the Choi-Jamio{\l}kowski isomorphism~\cite{zyczkowski}, which can be seen as an operational recipe for converting a process into a state. By inserting one half of a maximally entangled state $\ket{\psi^+} = \sum_{j} \ket{jj}/\sqrt{d}$ into the process described by \textsc{cptp} map $\Lambda$, a state $\Upsilon = \Lambda\otimes \mathcal{I} \left[\Psi^+ \right]$ (where $\Psi^+=\ket{\psi^+} \bra{\psi^+}$) can be constructed, whose matrix elements directly correspond to elements of $\Lambda$.

\subsection{Choi representation for multi-time processes}
\label{app:choi}

Here, we develop an analogue of the Choi-Jamio{\l}kowski isomorphism for more general processes.
Characterising the corresponding state is no easier than characterising the process tensor from the perspective of the number of parameters.
However, a range of techniques have been developed for efficient quantum state tomography~\cite{cramer, 10gross150401}. Owing to the isomorphism, such techniques are immediately available for quantum process tomography, rendering it efficient. Our claim is formalised in the following theorem:

\begin{theorem}
Any $k$-step process can be operationally represented by the generalised Choi state $\Upsilon_{k:0}$ of a $2k+1$--body system. $\Upsilon_{k:0}$ can further be written in matrix product operator form~\cite{MPS1}, with a bond dimension that is bounded by the effective dimension of the environment. \label{thm:CJI}
\end{theorem}

The generalised Choi state $\Upsilon_{k:0}$, corresponding to the process tensor $\mathcal{T}_{k:0}$, can be prepared experimentally using the circuit presented in Figure~\ref{fig:CJIcircuit}. We provide a detailed proof of this theorem in Appendix~\ref{app:CJIproof}, where we use the dilated OQE in Eq.~\eqref{eq:oqe} to demonstrate that the elements of the density operator that results from the circuit in Figure~\ref{fig:CJIcircuit} are exactly the elements of the corresponding process tensor. The action of the process tensor $\mathcal{T}_{k:0}$ on a set of operations $\mathbf{A}_{k-1:0}$ is equivalent to projecting the Choi state $\Upsilon_{k:0}$ onto the Choi state of $\mathbf{A}_{k-1:0}$ (up to a transpose), i.e., $\mathcal{T}_{k:0}[\mathbf{A}_{k-1:0}] = {\rm tr}_{S}[\Upsilon_{k:0}(\mathbbm{1}_{S}\otimes\mathcal{A}_{k-1}\otimes\mathcal{I}\otimes\dots\otimes\mathcal{A}_{0}\otimes\mathcal{I}[(\Psi^+)^{\otimes k-1}])]$, where the partial trace is over all subsystems except the one corresponding to the output of the process tensor (the system $S$ in Fig.~\ref{fig:CJIcircuit}).

\begin{figure}
\begin{center}
\includegraphics[width=.65 \linewidth]{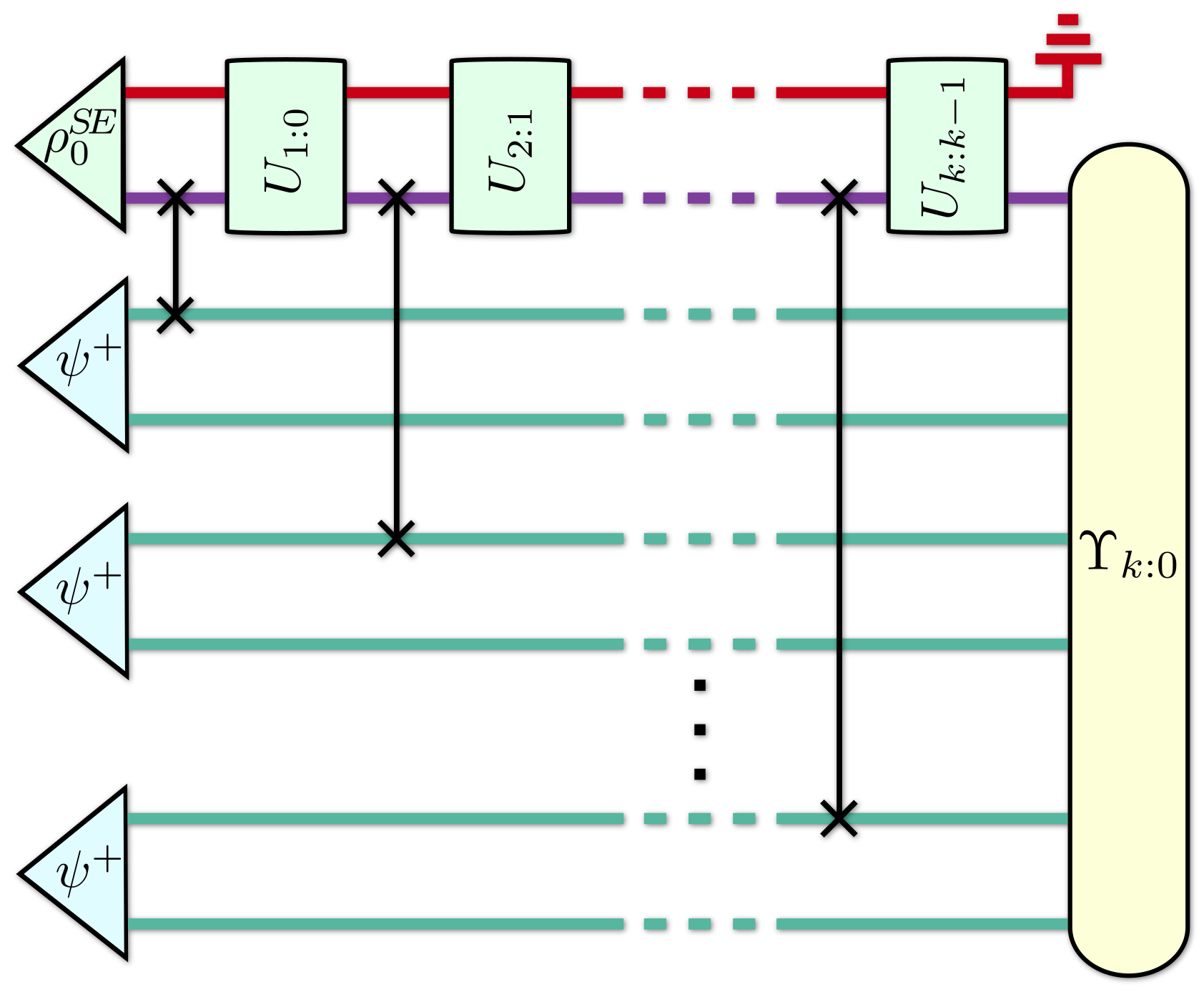}
\caption{\emph{Generalised Choi-Jamio{\l}kowski isomorphism.} This quantum circuit prepares the state that represents the process tensor element by element. The resources required are $k$ maximally entangled pairs of ancillas of dimension $d$. That is $k \, \log_2(d)$ ebits. Correlations between pairs of ancillas in $\Upsilon_{k:0}$ correspond directly to the memory inherent in the non-Markovian evolution. Any desired element of this state can be sampled using the techniques of quantum state tomography. See Appendix~\ref{app:CJIproof} for details. \label{fig:CJIcircuit}}
\end{center}
\vspace{-20pt}
\end{figure}

The Choi state representation allows direct access to important properties of the process, and maps temporal correlations onto spatial ones. Operational and mathematical tools, developed to understand, e.g. entanglement scaling, in many-body quantum states can now be applied directly to general dynamical quantum processes. This, in turn, enables the systematic classification and bounding of memory effects.

\subsection{Matrix product operator form}
\label{app:mps}

Our Theorem also implies that many physically relevant quantum processes will have an efficient description. In the second part of our proof, we show that the Choi state has a natural matrix product operator form; with the addition of two extra ancillas, the process tensor can be described by a pure, matrix-product state (MPS), which arises due to the causal nature of the process (this can also be seen in the second part of the proof of Theorem~\ref{thm:rep}). 

In short, we show that the Choi state for a given OQE can be written as the matrix product density operator~\cite{VerstraeteCirac2004}
\begin{align}\label{eq:MPSform}
\Upsilon_{k:0} =& \sum
M_k^{r_k r_{k-1}' s_k s_{k-1}'} \cdots M_1^{r_1 r_0' s_1 s_0'} M_0^{r_0 s_0} \\ \nonumber
& \quad \times \ket{r_k \,r'_{k-1} \dots r_1\, r'_0\, r_0}\bra{s_k\, s'_{k-1}\,  \dots s_1\, s'_0\, s_0},
\end{align}
composed of $d_{E}^2\times d_{E}^2$ matrices 
\begin{gather}
M_j^{r_j r_{j-1}' s_j s_{j-1}'} \!=\! \bra{r_j} \! U_{j:j-1} \! \ket{r'_{j-1}} \!\otimes\! \bra{s_j} \!{U^*_{j:j-1}}\! \ket{s'_{j-1}}\!,
\end{gather}
$j\neq 0,k$, and length $d_{E}^2$ row and column vectors 
\begin{gather}
\begin{split}
M_k^{r_k r_{k-1}' s_k s_{k-1}'} \!=\!& \sum_\epsilon \! \bra{r_k\epsilon} \!U_{k:k-1}\! \ket{r'_{k-1}} 
\\ & \quad \!\otimes\! \bra{s_k\epsilon} \!{U^*_{k:k-1}}\! \ket{s'_{k-1}}
\end{split}
\end{gather}
and
\begin{gather}
M_0^{r_0 s_0} = \sum_{\epsilon_0 \gamma_0}\rho^{SE}_{r_0 \epsilon_0; s_0 \gamma_0} \ket{\epsilon_0 \gamma_0}
\end{gather}
respectively---note that the superscripts here are \emph{not} matrix indices.

From this representation it is clear that the number of independent elements of the process tensor does not always grow exponentially with the number of time steps $k$, but will, in many physically relevant cases, grow linearly: the size of the matrix product operator and hence the process tensor grows as $\mathcal{O}(k d_{S}^2 D^2)$, where $D$ is the bond dimension of the state. In general, $D\le d_{E}$, the dimension of the environment. This is reassuring, since the description should not be more complex than the corresponding OQE. Even though the environment dimension could be large, there is always a consistent OQE with $d_{E}\leq d^{3^k-1}$, and we expect the effective bond dimension to be much smaller than this in practice; often only part of the environment interacts with the system at any given time and, in practice, even an infinite-dimensional environment can be approximated by a finite one~\cite{reactioncoord, generalpseudomodes}. This comprises a significantly more efficient representation for processes with many time steps, and opens up the possibility to use singular value truncation and other techniques~\cite{MPSreview, 10gross150401, cramer, arXiv:1612.08000} to meaningfully approximate the dynamics by pruning low-probability branches of the MPS description. We now demonstrate how the Choi state of a process tensor, defined on a set of time steps, can be used to directly recover information about dynamics on subsets of those time steps.

\subsection{Intermediate dynamics from the Choi state}
\label{app:subsys}

In a direct application of the containment property of the process tensor, we can recover the Choi state of smaller process tensors from $\Upsilon_{k:0}$. Taking the trace over the subsystem corresponding to the final output state ($S$ in Fig.~\ref{fig:CJIcircuit}) gives ${\rm tr}_{S}[\Upsilon_{k:0}] = \mathbbm{1}\otimes\Upsilon_{k-1:0}$; partial tracing over one further subsystem gives the Choi state of the process tensor up to $k-1$. By iterating this procedure and projecting other unwanted time steps on the maximally entangled state (corresponding to applying an identity operation), one can recover any intermediate process tensor. Specifically, if we split the time steps into a set we are interested in $j_1,j_2,\dots j_n$ and a set we wish to ignore $l_0,l_1,\dots l_{k-n}$ then we have:
\begin{gather}
\Upsilon_{j_n,\dots,j_2,j_1} = {\rm tr}_{j_n',\{l_i,l_i'\}}\left[\left(\Psi^+_{l_{k-n}}\otimes\dots\otimes\Psi^+_{l_{1}}\otimes\Psi^+_{l_{0}}\right) \Upsilon_{k:0}\right],
\end{gather}
where the primed and unprimed subsystem labels correspond to those in Eq.~\eqref{eq:MPSform} (depending on whether the final time step $k$ is included in the set $\{j_i\}$, some of the subscripts on the partial trace may be redundant). Replacing the maximally entangled states $\Psi^+$ in this equation with the Choi states of operations other than the identity will result in the Choi state of a conditional process tensor that corresponds to the case where those operations were applied. An important special case of this is where the $\Psi^+$ are replaced by identity operators $\mathbbm{1}$---the Choi state of a maximally incoherent operation (where all inputs are mapped to the maximally mixed state). $\Upsilon_{j_n,\dots,j_2,j_1}$ then simply becomes a reduced state of $\Upsilon_{k:0}$. In other words, the reduced states of $\Upsilon_{k:0}$ are Choi states of intermediate processes averaged over possible operations that may have been performed at other time steps.

We can also recover the Choi states of dynamical maps $\Lambda_{l:j}$ that take a freshly prepared system state at time step $j$ to that at later time step $l$:
\begin{align}
\Lambda_{l:j}\otimes\mathcal{I}[\Psi^+] = {\rm tr}_{\overline{jl'}}\left[\left(\bigotimes_{\tiny{m\neq j,l}}\! \Psi^+_m\right)\Upsilon_{k:0}\right];
\end{align}
here the trace is over all subsystems but those labelled by $j$ and $l'$.
In the case that $j=0$ and the system is uncorrelated with its environment initially, we recover the usual dynamical map as discussed in Sec.~\ref{app:conv}. These intermediate dynamical maps are always guaranteed to be completely positive, and are exactly what one would reconstruct if usual quantum process tomography were performed between step $j$ and step $l$ (assuming that the initial preparation procedure involves a deterministic entanglement-breaking operation).
Finally, when all but the subsystem corresponding to index $r_0$ in Eq.~\eqref{eq:MPSform} is traced out, we are merely left with the average initial system state ${\rm tr}_{\overline{0}}[\Upsilon_k] = {\rm tr}_{E}[\rho_0^{SE}] = \rho_0^{S}$.

\section{Discussion}

We have presented a universal framework for characterising arbitrary quantum processes, including non-Markovian ones, demonstrating that the process tensor is the most general descriptor of a quantum process. Our framework does not rely on any microscopic models; we only assume that experimental control operations act solely on ${S}$ and do not directly influence ${E}$ (in practice, this could be used as a definition for $S$). We have also shown how this characterisation could be made efficient by casting the process tensor as a  many-body state, with a matrix product operator form. Moreover, in our accompanying Letter~\cite{PRL_partner} we have used this new framework to derive consistent, unambiguous and meaningful measures of non-Markovianity, based on an operational condition for quantum Markov dynamics. Our methods could further be applied to continuous control by making use of the Trotter formula for the decomposition of the dynamics of a system.

By reconstructing, either numerically or in an experiment, the dynamics of a system in the form of a process tensor, an effective memory length and magnitude can be determined at a coarse-grained level by studying the correlations and bond dimension of the corresponding Choi state. Based on this, an effective description of the system could be constructed---using, for example, the transfer tensor method~\cite{CerrilloCao2014, PollockModi2017}---in the form of an approximate master equation. This simpler description would capture the essential features of a complex system's dynamics, while discarding those details which are superfluous at a given time scale.

This work further opens up many other avenues for future research. Apart from the possibility to derive non-Markovian master equations---by taking the limit of time-steps becoming infinitesimally close---the process tensor could be used to systematically study the properties of a typical process, analyse temporal quantum correlations~\cite{thao} and structures without causal order~\cite{PhysRevA.88.022318, OreshkovETAL2012}. On the practical side, it could be used for characterising electronic dynamics in molecules using spectroscopic techniques, or formalising adaptive quantum machine learning algorithms. Also, the \textsc{cp}~nature of the process tensor enables the calculation of its Holevo capacity~\cite{PhysRevA.92.052310}, which bounds the information content carried by a non-Markovian channel~\cite{wilde2013quantum}. Moreover, our approach paves the way for a general theory of non-Markovian error correction~\cite{11kalai11060485, 13preskill}.

Related representations for general quantum stochastic processes have appeared in the literature as early as 1979 and 1982~\cite{Lindblad1979,accardi}, albeit from a less operational starting point. More recently, the approach to modelling quantum channels with memory in Ref.~\cite{PhysRevA.72.062323} has led to a similar mathematical theory. In other contexts, the mathematical structure of the process tensor is also related to other formalisms which describe maps acting on quantum operations, notably the quantum combs~\cite{PhysRevLett.101.060401, PhysRevA.80.022339}, operator tensor~\cite{Hardy3385}, and process matrix~\cite{oreshkov2016, costashrapnel2016} frameworks. However, it has not hitherto been applied to the question of open quantum dynamics \emph{per se}; here we have constructed, for the first time, an operationally meaningful prescription to characterise an arbitrary open process across multiple time steps. The representation of the process as a matrix product operator also provides a novel tool for its efficient reconstruction.

\begin{acknowledgments}
We are grateful to A. Aspuru-Guzik, G. Cohen, A. Gilchrist, J. Goold, M. W. Hall, T. Le, K. Li, L. Mazzola, S. Milz, F. Sakuldee, D. Terno, S. Vinjanampathy, H. Wiseman, C. Wood, M.-H. Yung for valuable conversations. CR-R  is supported by MSCA-IF-EF-ST - QFluctTrans 706890. MP is supported by the EU FP7 grant TherMiQ (Grant Agreement 618074), the DfE-SFI Investigator Programme (grant 15/IA/2864), the H2020 Collaborative Project TEQ (grant 766900), and the Royal Society. KM is supported through ARC FT160100073.
\end{acknowledgments}

\bibliography{Pollockbib}

\onecolumngrid

\newpage
\appendix

\section{Index notation for quantum operations}

Throughout this article \textsc{cptp}~control operations $\mathcal{A}$ are taken to be Hermitian, positive, trace-$d$ matrices. In the remainder of the Appendices, we make extensive use of index notation, which we detail here. The action of the map $\mathcal{A}_j$ is defined as 
\begin{gather}
\mathcal{A}_{j}[\rho_j] = \sum_{r'_j,s'_j} \mathcal{A}_{r_jr'_j; s_js'_j}\rho_{r'_j;s'_j} \ket{r_j}\bra{s_j}.
\end{gather}
Note that index $j$ on the left denotes a time step. On the right we have expressed this as a subscript to matrix indices $r,r',s,s'$; these subscripts should not be interpreted merely as labels for dummy indices, but also reference the time step to which the indexed operator corresponds. Alternatively, we can write the action of the map in the Sudarshan-Kraus form as $\mathcal{A}(\rho) = \sum_l A_l \rho A_l^\dag$. See the `B-form' of the map in Ref.~\cite{sudarshan} for details of this representation of the map.

We write the action of a unitary transformation that takes the state of ${S\mbox{-}E}$ from time step $j$ to time step $k$ as
\begin{align}
\rho^{SE}_k =&\sum_{r_k\epsilon_k s_k\gamma_k} \rho^{SE}_{r_k\epsilon_k, s_k\gamma_k} \ket{r_k\epsilon_k} \bra{ s_k\gamma_k}\\
U_{k:j} \rho^{SE}_j U_{k:j}^\dagger =& \sum_{\substack{r_k\epsilon_ks_k\gamma_k \\ r_j\epsilon_j s_j\gamma_j}}
U_{r_k\epsilon_k,r_j\epsilon_j} \;
\rho^{SE}_{r_j\epsilon_j, s_j\gamma_j}
\; U^*_{s_k\gamma_k, s_j\gamma_j} \; \ket{r_k\epsilon_k} \bra{ s_k\gamma_k}.
\end{align}
Note that the input and the output indices of the unitary operators have different subscripts denoting the time steps they belong to. We rewrite the last equation as a quantum map as
\begin{gather}
\rho^{SE}_k = \mathcal{U}_{k:j}[\rho^{SE}_j]
=\sum_{\substack{r_k\epsilon_ks_k\gamma_k\\r_j\epsilon_j s_j\gamma_j}}
\mathcal{U}_{r_k \epsilon_k, r_j \epsilon_j; s_k \gamma_k, s_j \gamma_j} \;
\rho^{SE}_{r_j \epsilon_j, s_j \gamma_j}
\ket{r_k\epsilon_k} \bra{ s_k\gamma_k}.
\end{gather}
Finally, note that often we will omit the `kets' and `bras' from such equations.

As an example consider where unitary is acting on ${SE}$ and control operation is acting only on ${S}$:
\begin{align}
\rho^{S}_2 =&\mathcal{U}_{2:1} \, \mathcal{A}_1 \, \mathcal{U}_{1:0} \, \mathcal{A}_0 [\rho^{SE}_0] \nonumber\\
=&\sum_{\overline{r_2s_2}}
 \mathcal{U}_{r_{2} \epsilon_{2} x_1 \epsilon_1; s_{2} \gamma_{2} y_1 \epsilon_1} 
 \mathcal{A}_{x_1 r_1; y_1 s_1}  
\mathcal{U}_{r_{1} \epsilon_{1}, x_0 \epsilon_0; s_{1} \gamma_{1} y_0 \epsilon_0} 
 \mathcal{A}_{x_0 r_0; y_0 s_0}  
 \rho^{SE}_{r_0 \epsilon_0, s_0 \gamma_0}
 \ket{r_2} \bra{s_2} \nonumber\\
=&\sum_{\overline{r_2s_2}} \,
\prod_{j=0}^{1}
 \mathcal{U}_{r_{j+1} \epsilon_{j+1} x_j \epsilon_j; s_{j+1} \gamma_{j+1} y_j \epsilon_j} 
 \mathcal{A}_{x_j r_j; y_j s_j}  
 \rho^{SE}_{r_0 \epsilon_0, s_0 \gamma_0}
 \ket{r_2} \bra{s_2}.
\end{align}
Here, $\overline{x y}$ indicates that all but indices $x$ and $y$ should be summed over.

\section{Proof of representation theorem}

\subsection{Open quantum evolution implies process tensor}\label{app:proof:thm:qdptopt}

To prove the first part of Theorem~\ref{thm:rep} we need to derive the process tensor from the open quantum evolution in Eq.~\eqref{eq:oqe} and show that 
it satisfies the three properties prescribed below Definition~\ref{totalmap}. We begin by writing down Eq.~\eqref{eq:oqe} in terms of matrix indices. The state of the system at the time step $k$ is $\rho^{S}_k = {\rm tr}_{E}[ \rho^{SE}_k]$, and is a function of $\mathbf{A}_{k-1:0} = \{\mathcal{A}_{k-1}, \dots, \mathcal{A}_{0}\}$. We can write down this state in terms of matrix indices of these maps:
\begin{align} \label{eqindices}
\rho^{S}_{r_k,s_k} =& 
\sum_{\substack{r_{k-1} \cdots r_0 \\ x_{k-1} \cdots x_0}} \sum_{\substack{s_{k-1} \cdots s_0 \\ y_{k-1} \cdots y_0}} \sum_{\substack{\epsilon_k\cdots \epsilon_0 \\ \gamma_k\cdots \gamma_0}} \delta_{\epsilon_k\gamma_k} \prod_{j=0}^{k-1} \mathcal{U}_{r_{j+1} \epsilon_{j+1} x_j \epsilon_j; s_{j+1} \gamma_{j+1} y_j \gamma_j} \mathcal{A}_{x_j r_j; y_j s_j}  \rho^{SE}_{r_0 \epsilon_0, s_0 \gamma_0}\nonumber\\
=& \sum_{\substack{r_{k-1}\cdots r_0 \\ x_{k-1}\cdots x_0}}
\sum_{\substack{s_{k-1}\cdots s_0 \\ y_k\cdots y_0}}
\left(
\sum_{\substack{\epsilon_k\cdots \epsilon_0 \\ \gamma_k\cdots \gamma_0}}
\delta_{\epsilon_k \gamma_k}
\prod_{j=0}^{k-1}
\mathcal{U}_{r_{j+1} \epsilon_{j+1} x_j \epsilon_j; s_{j+1} \gamma_{j+1} y_j \gamma_j} 
\rho^{SE}_{r_0 \epsilon_0, s_0 \gamma_0}\! \right)\!\!
\left( \prod_{j=0}^{k-1}
\mathcal{A}_{x_j r_j; y_j s_j} \right)\nonumber\\
=& \sum_{\substack{r_{k-1}\cdots r_0 \\ x_{k-1}\cdots x_0}}
\sum_{\substack{s_{k-1}\cdots s_0 \\ y_k\cdots y_0}}
\mathcal{T}_{r_k, x_{k-1} r_{k-1} \cdots x_{0} r_{0};
s_k, y_{k-1} s_{k-1} \cdots y_{0} s_{0}}
\mathbf{A}_{x_{k-1} r_{k-1} \cdots x_{0} r_{0}; y_{k-1} s_{k-1} \cdots y_{0} s_{0}},
\end{align}
where the delta function in line one is the trace over the final state of ${E}$. In general, the initial state of the system can be correlated with the environment, which is not traced out until the final time step. Note that we have denoted the time-step indices as subscripts to matrix indices. Above, the process tensor and controls are defined as
\begin{align}\label{proctens}
\mathcal{T}_{r_k, x_{k-1} r_{k-1} \cdots x_{0} r_{0}; s_k, y_{k-1} s_{k-1} \cdots y_{0} s_{0}} =& \sum_{\substack{\epsilon_k\cdots \epsilon_0 \\ \gamma_k\cdots \gamma_0}} \delta_{\epsilon_k\gamma_k} \prod_{j=0}^{k-1} \mathcal{U}_{r_{j+1} \epsilon_{j+1} x_j \epsilon_j; s_{j+1} \gamma_{j+1} y_j \gamma_j} \rho^{SE}_{r_0 \epsilon_0, s_0 \gamma_0} \\
\mathbf{A}_{x_{k-1} r_{k-1} \cdots x_{0} r_{0}; y_{k-1} s_{k-1} \cdots y_{0} s_{0}} =& \prod_{j=0}^{k-1} \mathcal{A}_{x_j r_j; y_j s_j}.\label{control}
\end{align}
The element by element product in the last equation is simply a tensor product of operations $\mathcal{A}$ at different times. That is, the controls at different times are independent of each other. If these operations were correlated then we would have a more complex entity for Eq.~\eqref{control}.

\emph{Linearity of the process tensor} can be seen by substituting $\mathbf{A}_{\rm tot} = p \mathbf{A} +(1-p) \mathbf{B}$ into Eq.~\eqref{eqindices} and finding $\mathcal{T}[\mathbf{A}_{\rm tot}]= p \mathcal{T}[\mathbf{A}] +(1-p) \mathcal{T}[\mathbf{B}]$. We can interpret this linearity by considering a coin with probabilities $p$ and $1-p$ for `heads' and `tails' respectively. The coin is flipped and the outcome determines the choice of control operation, $\mathbf{A}^{(1)}$ or $\mathbf{A}^{(2)}$. Subsequently, the process outputs state $\mathcal{T}[\mathbf{A}^{(1)}]$ or $\mathcal{T}[\mathbf{A}^{(2)}]$. Interestingly, the value of $p$ or $1-p$ need not be positive (aforementioned example aside); the linearity condition holds for any linear expansion of controls $\mathbf{A}$, so long as their combination remains a valid set of operations.

\emph{Complete positivity for the process tensor} can be shown by casting it in the Sudarshan-Kraus-Choi form~\cite{sudarshan, kraus, choi72, choi75}; a linear map $\Lambda$ is \textsc{cp}~if and only if it can be decomposed as $\Lambda(\rho) = \sum_n L_n \rho L_n^\dag$. In our case, we make use of the matrix form of unitary operations, $\mathcal{U}[\rho]=U\rho U^\dagger$, to split their action from the left and right as:
\begin{align}\label{krausform}
\rho^{S}_{r_k,s_k} =& 
\sum_{\substack{r_{k-1}\cdots r_0 \\ x_{k-1}\cdots x_0}}
\sum_{\substack{s_{k-1}\cdots s_0 \\ y_k\cdots y_0}}
\sum_{\substack{\epsilon_k \cdots \epsilon_0 \\ \gamma_k\cdots \gamma_0}}
\delta_{\epsilon_k \gamma_k}
\left(\prod_{j=0}^{k-1}
U_{r_{j+1} \epsilon_{j+1} x_j \epsilon_j}
\sqrt{\rho^{SE}_{r_0 \epsilon_0, s_0 \gamma_0}} \right) \nonumber \\ 
&\qquad\qquad\qquad\qquad\qquad\qquad\times
\prod_{j=0}^{k-1} \mathcal{A}_{x_j r_j; y_j s_j}
\left(\prod_{j=0}^{k-1}
\sqrt{\rho^{SE}_{r_0 \epsilon_0, s_0 \gamma_0}}
U^*_{s_{j+1} \gamma_{j+1} y_j \gamma_j} 
 \right) \nonumber \\
=&\!\! \sum_{\substack{r_{k-1}\cdots r_0 \\ x_{k-1}\cdots x_0}}
\sum_{\substack{s_{k-1}\cdots s_0 \\ y_k\cdots y_0}}
\!\!(T_l)_{r_k, x_{k-1} r_{k-1} \cdots x_{0} r_{0}}
\mathbf{A}_{x_{k-1} r_{k-1} \cdots x_{0} r_{0}; y_{k-1} s_{k-1} \cdots y_{0} s_{0}}
(T_l)^*_{s_k, y_{k-1} s_{k-1} \cdots y_{0} s_{0}},
\end{align}
where we have used the positivity of the initial state to take its square root. We have achieved the desired form and thus proven complete positivity. From Eq.~\eqref{krausform}, we can write the operators $T_l$ in Eq.~\eqref{eq:cpform} of the main text as 
\begin{align}
\left(T_{\epsilon_k \cdots \epsilon_0 \gamma_k\cdots \gamma_0} \right)_{r_k, x_{k-1} r_{k-1} \cdots x_{0} r_{0}} = 
\prod_{j=0}^{k-1}
U_{r_{j+1} \epsilon_{j+1} x_j \epsilon_j}
\sqrt{\rho^{SE}_{r_0 \epsilon_0, s_0 \gamma_0}},
\end{align}
where $l = \epsilon_k \cdots \epsilon_0 \gamma_k \cdots \gamma_0$.

\emph{Containment property of the process tensor}--- implying $\mathcal{T}_{k:j}$ contains $\mathcal{T}_{k':j'}$ for $j \le j' \le k' \le k$---can be seen by letting all controls from $j$ to $j'$ be the identity map. This yields the total ${S\mbox{-}E}$ state  $\rho^{SE}_{j'}$. Next, we allow arbitrary controls from $j'$ to $k'$ and then discontinue the evolution. This is just a special case of the procedure above with specific choices of controls outside of the interval $[j',k']$. However, within the interval,  $\mathcal{T}_{k':j'}$ is fully constructed.  \hfill $\blacksquare$

\subsection{Proof that process tensor implies open quantum evolution}
\label{app:dilationproof}

The converse statement is a generalisation of the Stinespring dilation theorem~\cite{Stinespring}. In order to prove that all process tensors have a unitary representation, we first consider that, for a single time-step process,  $\mathcal{T}_{1:0}[\mathcal{A}_0] = \rho_1 = \$(\mathcal{A}_0)[\rho_0]$, where $\rho_0$ is some initial reduced state of the system and $\$$ is a \emph{supermap}~\cite{supermaps}, which maps operations on the system to other operations: $\$(\mathcal{A}_0)[\rho]=\mathcal{A}_0'[\rho]$. This description is possible due to the \textsc{cp}~nature of the process tensor and its resulting Kraus decomposition (see Sec.~\ref{app:proof:thm:qdptopt}). 

In Theorem 1 of Ref.~\cite{supermaps} it is proven that the action of a supermap can always be represented as
\begin{gather}
\$(\mathcal{A}_0)[\rho] = {\rm tr}_{A_0}\left\{W (\mathcal{A}_0\otimes\mathcal{I}_B)\left[Z \rho Z^\dagger\right] W^\dagger\right\},\label{eq:supermapdilation}
\end{gather}
where $Z : {S}\to {S}\times {B}_0$ and $W : {S} \times {B}_0\to {S}\times {A}_0$ are isometries acting on the system and two ancillas ${A}_0$ and ${B}_0$, and we have written the identity map on the ancilla explicitly. Since the processes we are considering do not change the dimension of the system, we can take ${A}_0$ and ${B}_0$ to be the same and of dimension $d_{{A}_0}\ge d^2$. In this case $W$ corresponds to a unitary map $\mathcal{W}$ on the joint system-ancilla space. Moreover, we can rewrite $Z \rho Z^\dagger = \mathcal{V}\, \left[\rho \otimes \eta_{0} \right]$, where $\mathcal{V}$ is another unitary map on the system-ancilla space and $\eta_0$ is the initial state of the ancilla. Therefore, we have
\begin{gather}
\mathcal{T}_{1:0}[\mathcal{A}_0] = \$(\mathcal{A}_0)[\rho_0] = {\rm tr}_{A_0}\left\{\mathcal{W}_{1:0} \mathcal{A}_0\mathcal{V}_{0}\left[ \rho_0 \otimes \eta_{0} \right] \right\}. \label{eq:superchanneldilation}
\end{gather}
Here, $\mathcal{A}_0$ acts on the system alone; there is an implied identity map on the ancilla.

Let's assume that for the process up to step $j-1$, $\mathcal{T}_{j-1:0}[\mathbf{A}_{j-2;0}]$ can be represented by unitary evolution of the form 
\begin{align}
\mathcal{T}_{j-1:0}[\mathbf{A}_{j-2;0}] =& {\rm tr}_{{E}_{j-1}}\left\{ \, \mathcal{W}_{j-1:j-2} \, \mathcal{A}_{j-2} \, \mathcal{V}_{j-2}
\left[\dots 
\left[ \, \mathcal{W}_{1:0} \, \mathcal{A}_0 \, \mathcal{V}_{0} \left[\rho_0\otimes\eta_0 \right]\,\otimes\eta_1\right]\dots \,\otimes\eta_{j-2} \right]  \right\} \nonumber\\
=& {\rm tr}_{{E}_{j-1}}\left\{\rho^{{SE}}_{j-1}\right\}, \label{eq:dilationtheorem}
\end{align}
where ${E}_{j-1}$ is the environment consisting of ancillas $A_0$ to $A_{j-2}$. An additional step, with operation $\mathcal{A}_{j-1}$, can be added to the process by considering the evolution as another supermap $\$_j$ acting on the joint operation $\mathcal{A}_{j-1}\otimes \mathcal{I}_{{E}_{j-1}}$, the result of which then acts on the state $\rho^{{SE}}_{j-1}$. In other words, $\mathcal{T}_{j:0}[\mathbf{A}_{j-1;0}] = {\rm tr}_{{E}_{j-1}}\{\$_j(\mathcal{A}_{j-1}\otimes \mathcal{I}_{{E}_{j-1}}) [\rho^{{SE}}_{j-1}]\}$. We can then use Eq.~\eqref{eq:supermapdilation} to write
\begin{align}
\mathcal{T}_{j:0}[\mathbf{A}_{j-1;0}] = {\rm tr}_{{E}_j}\left\{\mathcal{W}_{j:j-1} \mathcal{A}_{j-1}\mathcal{V}_{j-1}\left[ \rho^{{SE}}_{j-1} \otimes \eta_{j-1} \right] \right\},\label{eq:inducted}
\end{align}
where the new ancilla has dimension $d_{A_{j}}\ge d^{2(3^j)}$. This evolution is of the same form as Eq.~\eqref{eq:dilationtheorem}, thus by induction it follows from Eqs.~\eqref{eq:superchanneldilation}~\&~\eqref{eq:inducted} that the evolution in Eq.~\eqref{eq:dilationtheorem} is valid for any time step. 

If we define $\rho_0^{SE} = \mathcal{V}_0[\rho_0 \otimes \eta_0] 
\otimes \eta_1 \dots \otimes \eta_{k-1}$ as the initial system-environment state, then the process tensor is consistent with OQE as defined in Eq.~\eqref{eq:oqe} with $\mathcal{U}_{j:j-1} = \mathcal{V}_j\mathcal{W}_{j:j-1}$ for $j<k$ and $\mathcal{U}_{k:k-1} = \mathcal{W}_{k:k-1}$. \hfill $\blacksquare$

\section{Proof that the process tensor has a matrix-product state representation (Theorem~\ref{thm:CJI})}
\label{app:CJIproof}
In this proof, we make use of Theorem~\ref{thm:dilation} to represent the process tensor as an OQE with some environment; we further introduce a set of $2k$ ancillas each of $d$-dimensions, which along with the system, will be used to encode the many-body state. This theorem generalises the well-known Choi-Jamio{\l}kowski
isomorphism~\cite{jamiolkowski} to the process tensor.

Let us label the pair of ancillas to be used at the $j$th time step ${A}_j$ and ${B}_j$, these are initialised in the maximally entangled state $\ket{\psi^+}_{ {A}_j {B}_j} = \sum_{x_j=1}^d \ket{x_j x_j}/{\sqrt{d}}$.  Let the total state of  system--environment--ancillas at time step $j$ be
\begin{align}
\Theta_{j} =& \sum
\Theta_{r'_{j} \epsilon_{j} x_{j-1} y_{j-1} \cdots x_{1} y_{1} x_{0} y_{0}, s_{j}'\gamma_{j} w_{j-1} z_{j-1} \cdots w_{1} z_{1} w_{0} z_{0}} 
\\ \notag  & \hspace{1cm} \times
\ket{r'_{j} \epsilon_{j} x_{j-1} y_{j-1} \dots x_{1} y_{1} x_{0} y_{0}}
\bra{s'_{j} \gamma_{j} w_{j-1} z_{j-1} \dots w_{1} z_{1} w_{0} z_{0}}
\end{align}
Above the indicies $\{r'_{j}, s'_{j} \}$ \& $\{ \epsilon_{j}, \gamma_{j} \}$ belong to ${S}$ \& ${E}$ respectively, and $\{x_{l}, w_{l} \}$ \& $\{ y_{l}, z_{l} \}$ belong ancillas ${A}_l$ and ${B}_l$ respectively with $0 \le l \le j-1$. In each case the subscript on the index denotes the time step. Thus $\Theta_{j}$ includes ancillas $\{{A}_{j-1} {B}_{j-1} \dots {A}_{0} {B}_{0} \}$. Next we apply the \textsc{swap} operation $\mathbf{S}_j$ to ${S}$ and ancilla ${A}_j$, defined as $\mathbf{S} \ket{rx} = \ket{xr}$. This gives us
\begin{align}
\mathbf{S}_{j}^{{SA}} 
\Theta_{j} \otimes \ket{\psi^+}_{{A}_{j}{B}_{j}} \bra{\psi^+}  \mathbf{S}_{j}^{{SA}} 
= \frac{1}{d} \sum \Theta_{r'_{j} \epsilon_{j} x_{j-1} y_{j-1} \cdots x_{1} y_{1} x_{0} y_{0}, s_{j}'\gamma_{j} w_{j} z_{j} \cdots w_{1} z_{1} w_{0} z_{0}} \qquad&
\\ \notag  \times\ket{x_{j} \epsilon_{j} r'_j x_j} \bra{y_{j} \gamma_{j} s'_j y_j}\otimes
\ket{x_{j-1} y_{j-1} \dots x_{1} y_{1} x_{0} y_{0}}
\bra{w_{j-1} z_{j-1} \dots w_{1} z_{1} w_{0} z_{0}}.&
\end{align}
In the last equation the first line contains ${S}_j{E}_j{A}_j{B}_j$ and the second line contains the previous ancillas ${A}_{j-1} {B}_{j-1} \cdots {A}_0 {B}_0$. After the \textsc{swap} gate is applied the state is evolved to the next time step by the unitary map $\mathcal{U}_{j+1:j}$. The action of the unitary can be written
\begin{align}
\mathcal{U}_{j+1:j} \left(\ket{x_j\epsilon_j}\bra{y_j\gamma_j}\right) =& \sum
\mathcal{U}_{r_{j+1} \epsilon_{j+1},r_{j} \epsilon_{j} ; s_{j+1} \gamma_{j+1}, s_{j} \gamma_{j}} 
\nonumber\\
&\qquad\qquad\times\ket{r_{j+1} \epsilon_{j+1}}\bra{r_{j} \epsilon_{j}}
\left(\ket{x_j\epsilon_j}\bra{y_j\gamma_j} \right)
\ket{s_{j} \gamma_{j}}\bra{s_{j+1} \gamma_{j+1}} \\
=& \sum
\mathcal{U}_{r_{j+1} \epsilon_{j+1},r_{j} \epsilon_{j} ; s_{j+1} \gamma_{j+1}, s_{j} \gamma_{j}} 
\ket{r_{j+1} \epsilon_{j+1}}\bra{s_{j+1} \gamma_{j+1}} \delta_{x_jr_j}\delta_{y_js_j}.
\end{align}
Combining these equations, the total system--environment--ancilla state at the next time step is
\begin{align}
\Theta_{j+1} = & \, \mathcal{U}_{j+1:j} \mathbf{S}_{j}^{{SA}} \, \Theta_{j} \otimes \ket{\psi^+}_{{A}_{j} {B}_{j}} \bra{\psi^+} \, \mathbf{S}_{j}^{{SA}} \\
=& \frac{1}{d} \sum
\mathcal{U}_{r_{j+1} \epsilon_{j+1},r_{j} \epsilon_{j} ; s_{j+1} \gamma_{j+1}, s_{j} \gamma_{j}} 
\Theta_{r'_{j} \epsilon_{j} x_{j} y_{j} \cdots x_{1} y_{1} x_{0} y_{0}, s_{j}'\gamma_{j} w_{j} z_{j} \cdots w_{1} z_{1} w_{0} z_{0}}
\label{eq:swapaction}  \\& \hspace{1cm} \notag 
\times \ket{r_{j+1} \epsilon_{j+1} r'_j r_j} \bra{s_{j+1} \gamma_{j+1} s'_j s_j}\otimes \ket{x_{j-1} y_{j-1} \dots x_{1} y_{1} x_{0} y_{0}}
\bra{w_{j-1} z_{j-1} \dots w_{1} z_{1} w_{0} z_{0}}.
\end{align}
Iterating Eq.~\eqref{eq:swapaction} with $\Theta_{0}=\rho_0^{SE}$ and taking the trace over the environment, we find for a $k$-step process
\begin{align}
\Upsilon_k = & {\rm tr}_{E}[\Theta_k] \nonumber \\
= & \frac{1}{d^{k}} \sum \delta_{\epsilon_k \gamma_k} \,
\mathcal{U}_{r_k \epsilon_k r'_{k-1} \epsilon_{k-1} ; s_k  \gamma_k s'_{k-1} \gamma_{k-1}}
\dots
\mathcal{U}_{r_{2} \epsilon_{2} r'_{1} \epsilon_{1} ; s_{2}  \gamma_{2} s'_{1} \gamma_{1}}
\mathcal{U}_{r_{1} \epsilon_{1} r'_{0} \epsilon_{0} ; s_{1}  \gamma_{1} s'_{0} \gamma_{0}}
\; \rho^{{SE}}_{r_0  \epsilon_0; s_0 \gamma_0} 
\notag\\ & \hspace{1cm} \times \ket{r_k r'_{k-1} r_{k-1} \dots r'_1 r_1 r'_0 r_0} \bra{s_k s'_{k-1} s_{k-1}  \dots s'_1 s_1 s'_0 s_0} \nonumber \\
= & \frac{1}{d^{k}} \sum \mathcal{T}_{r_k \,r'_{k-1} \cdots r_1\,r'_0\, r_0 ; s_k\, s'_{k-1} \cdots s_1\,s'_0\,s_0} \ket{r_k r'_{k-1} r_{k-1} \dots r'_1 r_1 r'_0 r_0} \bra{s_k s'_{k-1} s_{k-1}  \dots s'_1 s_1 s'_0 s_0}. \label{eq:CJIstate}
\end{align}
This is clearly a density operator with matrix elements corresponding to the components of the process tensor.

To prove that the state in Eq.~\eqref{eq:CJIstate} corresponds to a matrix-product state, we first realise that we can rewrite it as the matrix product density operator~\cite{VerstraeteCirac2004}
\begin{gather}
\Upsilon_k = \sum
M_k
^{r_k r_{k-1}' s_k s_{k-1}'}\cdots M_1^{r_1 r_0' s_1 s_0'} M_0^{r_0 s_0} \ket{r_k \,r'_{k-1} \dots r_1\, r'_0\, r_0}\bra{s_k\, s'_{k-1}\,  \dots s_1\, s'_0\, s_0},
\end{gather}
composed of $d_{E}^2\times d_{E}^2$ matrices 
\begin{gather}
M_j^{r_j r_{j-1}' s_j s_{j-1}'} = \bra{r_j}U_{j:j-1}\ket{r'_{j-1}}\otimes \bra{s_j}{U^*_{j:j-1}}\ket{s'_{j-1}},
\end{gather}
$j\neq 0,k$, and length $d_{E}^2$ row and column vectors 
\begin{gather}
M_k^{r_k r_{k-1}' s_k s_{k-1}'} = \sum_\epsilon \bra{r_k\epsilon}U_{k:k-1}\ket{r'_{k-1}}\otimes \bra{s_k\epsilon}{U^*_{k:k-1}}\ket{s'_{k-1}}
\end{gather}
and
\begin{gather}
M_0^{r_0 s_0} = \sum_{\epsilon_0 \gamma_0}\rho^{SE}_{r_0 \epsilon_0; s_0 \gamma_0} \ket{\epsilon_0 \gamma_0}
\end{gather}
respectively---note that the superscripts here are \emph{not} matrix indices. Given a decomposition of the initial state $\rho_{0}^{SE} = \sum_\lambda p_\lambda \ket{\phi_\lambda}\bra{\phi_\lambda}$, then the latter vector can be rewritten as $M_0^{r_0 s_0} = \sum_\lambda p_\lambda \amplitude{r_0}{\phi_\lambda} \otimes (\amplitude{s_0}{\phi_\lambda})^*$.

Aside from the subsystems corresponding to the initial and final time steps, the state is pure. It can thus be represented as a (pure) matrix-product state with only two ancillas, using the results of Ref~\cite{VerstraeteCirac2004}:
\begin{align}
\ket{\psi_{\mathcal{T}_k}} =& \sum
p_\lambda \bra{r_k\epsilon} U_{k:k-1} \ket{r'_{k-1}} \bra{r_{k-1}}U_{k-1:k-2} \ket{r'_{k-2}} \nonumber\\
&\qquad\qquad\qquad\qquad\dots  \bra{r_1}U_{1:0} \ket{r'_0}  \amplitude{r_0}{\phi_\lambda} \ket{r_k r'_{k-1} \dots r_1 r'_0 r_0 \epsilon \, \lambda}, \label{eq:MPS}
\end{align}
which has bond dimension $D = d_{E}$. 
However, as can be seen from the construction presented in Fig.~\ref{fig:dilateddynamics}, the minimal dimension of the environment, and hence the bond dimension, for the most general process grows with the number of time steps; this leads to a tree-like structure of the MPS description. This completes the proof. \hfill$\blacksquare$

\end{document}